\documentclass{aastex6}
\usepackage{graphicx}

\fullcollaborationName{The Friends of AASTeX Collaboration}

\begin{document}


\title{Spiral Arms, Infall, and Misalignment of the Circumbinary Disk
from the Circumstellar Disks in the Protostellar Binary System L1551 NE}



\author{Shigehisa Takakuwa\altaffilmark{1,2},
Kazuya Saigo\altaffilmark{3}, Tomoaki Matsumoto\altaffilmark{4},
Masao Saito\altaffilmark{5},
Jeremy Lim\altaffilmark{6}, Tomoyuki Hanawa\altaffilmark{7}, Hsi-Wei Yen\altaffilmark{2,8},
\& Paul T. P. Ho\altaffilmark{2,9}}

\altaffiltext{1}{Department of Physics and Astronomy, Graduate School of Science and Engineering,
Kagoshima University, 1-21-35 Korimoto, Kagoshima, Kagoshima 890-0065, Japan;
takakuwa@sci.kagoshima-u.ac.jp}
\altaffiltext{2}{Academia Sinica Institute of Astronomy and Astrophysics,
P.O. Box 23-141, Taipei 10617, Taiwan}
\altaffiltext{3}{ALMA Project Office, National Astronomical Observatory of Japan, Osawa 2-21-1,
Mitaka, Tokyo 181-8588, Japan}
\altaffiltext{4}{Faculty of Humanity and Environment, Hosei University, Chiyoda-ku, Tokyo 102-8160}
\altaffiltext{5}{Nobeyama Radio Observatory, National Astronomical Observatory of Japan,
Minamimaki, Minamisaku, Nagano 384-1805, Japan}
\altaffiltext{6}{Department of Physics, University of Hong Kong, Pokfulam Road, Hong Kong}
\altaffiltext{7}{Center for Frontier Science, Chiba University, Inage-ku, Chiba 263-8522, Japan}
\altaffiltext{8}{European Southern Observatory, Karl-Schwarzschild-Str. 2, Garching 85748, Germany}
\altaffiltext{9}{East Asian Observatory, 660 N. A'ohoku Place, University Park, Hilo, Hawaii 96720, U.S.A.}


\begin{abstract}
We report the ALMA Cycle 2 observations of the Class I binary
protostellar system L1551 NE in the 0.9-mm continuum,
C$^{18}$O (3-2), $^{13}$CO (3--2), SO (7$_8$-6$_7$),
and the CS (7--6) emission.
At 0$\farcs$18 (= 25 AU) resolution, $\sim$4-times higher than that of our Cycle 0 observations,
the circumbinary disk as seen in the 0.9-mm emission is shown
to be comprised of a northern and a southern spiral arm, with the southern arm
connecting to the circumstellar disk around Source B.
The western parts of the spiral arms are brighter than the
eastern parts, suggesting the presence of an $m=1$ spiral mode.
In the C$^{18}$O emission, the infall gas motions
in the inter-arm regions and the outward gas motions in the arms are identified.
These observed features are well reproduced with our numerical simulations,
where gravitational torques from the binary system impart angular momenta
to the spiral-arm regions and extract angular momenta from the inter-arm regions.
Chemical differentiation of the circumbinary disk is seen in the four molecular species.
Our Cycle 2 observations have also resolved the circumstellar disks around
the individual protostars, and the beam-deconvolved sizes are
0$\farcs$29 $\times$ 0$\farcs$19 (= 40 $\times$ 26 AU) (P.A. = 144$\degr$) and
0$\farcs$26 $\times$ 0$\farcs$20 (= 36 $\times$ 27 AU) (P.A. = 147$\degr$)
for Sources A and B, respectively. The position and inclination angles of these circumstellar disks
are misaligned with that of the circumbinary disk.
The C$^{18}$O emission traces the Keplerian rotation of the misaligned disk
around Source A.
\end{abstract}




\section{Introduction} \label{sec:intro}

Binary formation is likely a primary mode of solar-type star formation, since most ($\gtrsim$50$\%$)
solar-type stars \cite{du91b,rag10} and protostellar sources \cite{mur13,che13,rei14,tob16}
are members of binaries.
Protostellar binaries have been identified as pairs of circumstellar disks (hereafter CSDs)
surrounding each protostar \cite{loo00,lim06,mau10,tob15,tob16,lim16a,lim16b}.
In a number of protostellar binaries, circumbinary disks (hereafter CBDs) surrounding both of the CSDs
have also been identified \cite{tak04,tak12,tob13,cho14,tan14,tan16,dut14,dut16}.
Since CBDs can act as mass reservoirs to the protostellar binaries,
observational studies of internal structures and gas motions in CBDs
are crucial to understand how protostellar binaries grow, and how their final masses and mass ratios are determined.

In CBDs, tidal torques by orbiting protostellar binaries can clear the material
and create gaps within the disks. At the same time, the non-axisymmetric gravitational potential from the
binary (i.e., Roche potential), induces both inward and outward gas motions with slower and
faster rotations respectively, as compared to the standard Keplerian rotations. This results in spiral density
patterns in the CBDs. Previous theoretical studies of CBDs
around protostellar binaries predict such spiral-arm structures and non-axisymmetric
rotating and infalling gas motions \cite{bat00,gun02,och05,han10,dem15,you15}.
Observational identifications of such spiral-arm features and infalling motions
toward the protostellar binaries in the CBDs, are essential to understand the growth mechanism
of protostellar binaries. It is also important to measure the gas distribution
and gas motions between the CBD and CSDs, in order to understand
the transfer of material from the CBD to the CSDs.

Our group has been conducting a series of SMA, ASTE, and ALMA studies of
the Class I protostellar binary
L1551 NE, located in the L1551 region at a distance $d$=140 pc
\cite{tak11,tak12,tak13,tak14,tak15}. The source bolometric temperature and luminosity
are $T_{bol}$ = 91 K and $L_{bol}$ = 4.2 $L_{\odot}$ \cite{fro05}.
The projected binary separation and the position angle are
70 AU and 120$\degr$. The south-eastern source is referred to as
``Source A", and the north-western source as ``Source B" \cite{rei00,rei02,lim16b}. Our previous SMA
observations of L1551 NE have identified a $r$$\sim$300 AU-scale CBD, where the gas motions
can be modeled satisfactorily with a circular Keplerian rotation
with the central stellar mass of 0.8 $M_{\odot}$ \cite{tak12}.
The Keplerian CBD is surrounded by a $\sim$1000 AU-scale infalling envelope \cite{tak13}.
A larger scale ($\sim$20000 AU) molecular envelope is also seen,
which appears to be dispersed by the interaction with
the redshifted outflow driven from the neighboring protostar L1551 IRS 5 \cite{tak15}.

In the 0.9-mm dust-continuum emission, our ALMA Cycle 0 observations of L1551 NE at a $\sim$0$\farcs$7 resolution, resolved the structure of the CBD
and separated the CBD and the two CSDs.
In the C$^{18}$O (3--2) emission, deviations of
the gas motion from the Keplerian rotation in the CBD are also identified.
To compare with our data, we produced model ALMA Cycle 0 images from our hydrodynamic simulations.
The simulated images show
that the observed CBD structures can be interpreted as two spiral arms,
and that the observed gas motions can be interpreted as expanding and infalling
gas motions in the arm and inter-arm regions, respectively. This is because
gravitational torques from the central binary system impart angular momenta
to the spiral-arm regions and extract angular momenta from the inter-arm regions.

However, the Cycle 0 resolution was still too coarse
to fully characterize the spiral-arm structures, the gas motions within the spiral-arm and inter-arm
regions, and within the circumstellar disks.
Here, we present our
ALMA Cycle 2 observations of L1551 NE at a $\sim$4-times higher spatial resolution,
as compared to the Cycle 0 observations.
In Section 2 of the present paper, we describe our experiment, the data reduction and imaging,
and our numerical simulations
to be compared with the observed features. In Section 3, we present the results on the 0.9-mm
dust-continuum and C$^{18}$O ($J$=3--2) emission. We have
successfully resolved the spiral structures of the CBD and identified gas motions in the
spiral-arm and inter-arm regions. We have also resolved the structure of the individual
CSDs and identified the gas motions in the CSDs.
In Section 4, we present physical
interpretations of the observed features,
and the related issues of studies of protostellar binaries.
Section 5 provides a concise summary of our main results and discussion.

\section{Methods} \label{sec:met}
\subsection{ALMA Observations} \label{subsec:obs}

Our ALMA observations of L1551 NE were made on 2015 June 27.
Four molecular lines in Band 7, that is, the
C$^{18}$O ($J$=3-2; 329.3305453 GHz), the $^{13}$CO ($J$=3-2; 330.587965 GHz),
the SO ($J_N$=7$_8$-6$_7$; 340.71416 GHz), and the CS ($J$=7-6; 342.882857 GHz) lines,
as well as the 0.9-mm dust-continuum emission, were
observed simultaneously.
The precipitable water in the atmosphere 
was $\sim$0.55 -- 0.61 mm, an excellent condition during the Band 7 observations.
Excluding overheads for calibration, the total time on source was 34 minutes.
Table \ref{obs} summarizes the observational parameters.
The number of the available antennae was 36, but due to the correlator problem
it was 35 for the SO and CS lines as well as the corresponding continuum.
The C34-7/(6) configuration was adopted, which provides the minimum and maximum baseline lengths
of 33.7 m and 1.6 km, respectively. The projected baseline length on the source
ranges from 29 m to 1575 m. The minimum projected baseline length implies
that the present ALMA observation can recover $\sim$50$\%$
of the peak flux for a Gaussian
emission distribution with a FWHM of $\sim$2$\farcs$8 ($\sim$390 AU),
and $\sim$10$\%$ with a FWHM of $\sim$5$\farcs$2 ($\sim$730 AU) \cite{wil94}.
Thus, the present ALMA observation is sensitive to the structures of
the CBD with its outermost radius of $\sim$300 AU ($\sim$2$\farcs$1) around L1551 NE,
while largely insensitive to its surrounding envelope structures
with the outermost extent of $>$20000 AU ($\sim$143$\arcsec$) \cite{tak15}.



In its Frequency Division Mode (FDM), the ALMA correlator was configured
to provide four independent spectral windows (Basebands), each having a bandwidth of 468.75 MHz
and one of the four observed molecular lines.
The spectral window for the C$^{18}$O line
was divided into 3840 channels, with each channel having a width of 122.07 kHz.
Hanning smoothing was applied to the spectral channels,
resulting in a frequency resolution of 244.14 kHz and hence
a velocity resolution of 0.22 km s$^{-1}$ for the C$^{18}$O line.
Due to the limitation of the incoming data rate, 
the number of spectral channels and hence the
frequency resolution, had to be degraded
by a factor of 2 and 4 for the other spectral windows.
The resultant velocity resolutions are 0.44 km s$^{-1}$, 0.86 km s$^{-1}$,
and 0.85 km s$^{-1}$ for the $^{13}$CO, SO, and the CS lines, respectively.
In the following, we will use the higher-resolution C$^{18}$O line to discuss the velocity structures
of the CBD and CSDs in L1551 NE.
Channels in all four spectral windows, which are devoid of line emission,
were used to create the continuum image, which has
a central frequency of 335.85 GHz (= 0.893 mm) and a total bandwidth of 1.845 GHz.


Calibration of the raw visibility data was performed by the ALMA observatory
through the pipeline using
the Common Astronomy Software Applications (CASA) version 4.2.2. The quasar J0510+180, with estimated flux of $\sim$2.4 Jy, was
adopted as the gain and absolute flux calibrators. The quasar J0423-0120 was adopted
as the passband calibrator. We checked the tables of the pipeline calibration
as well as the calibrated visibility data. We confirmed that the calibration was fine
except for the presence of one integration of L1551 NE with unreasonably high amplitudes,
which we flagged manually before the imaging process. We did not perform self-calibration,
since the signal-to-noise ratio of the present ALMA data is high.
The calibrated visibility data were then Fourier-transformed
and CLEANed to create the continuum and molecular-line images. For the molecular line
imaging, Natural weighting of the visibility data was adopted
to enhance sensitivity. For the continuum imaging, various weighting of the visibility data were tried. 
The continuum image with the Natural weighting
is used to investigate the structures of the CBD.  The continuum images with uniform weighting, and with restriction of baselines longer than 450 m, resulted in higher
resolutions, were used to resolve and study the structures of the CSDs.
Corresponding angular resolutions are listed in Table \ref{obs}.

\subsection{Numerical Simulations} \label{subsec:sim}

To interpret the structures and kinematics of the CBD observed
with the present ALMA observation, we adopt the hydrodynamical numerical simulation
using adaptive mesh refinement (AMR) code SFUMATO \cite{mat07}. These simulations 
had reproduced very well the previous results of the ALMA Cycle 0 observation,
as described by Takakuwa et al. (2014).
We adopted the same model parameters as those in Takakuwa et al. (2014),
except for the orbital time in the simulation, in order to extract the
simulation data and to create the model images.
As described below, in our new Cycle 2 image of the 0.9-mm continuum emission,
it is clear that the western part is much brighter than the eastern part.
Such a one-sided, non-point-symmetric distribution of materials, is regarded
as an $m=1$ mode, which is distinct from an $m=2$ mode of two identical spiral arms.
Our numerical simulation shows that
after $\sim$20 orbital periods such an $m=1$ mode is well-developed
and is present continuously until the end of the simulation ($\sim$250 orbital periods).
This $m=1$ mode has not been explicitly reported
in the previous theoretical simulations of CBDs \cite{bat00,gun02,och05,han10,dem15,you15},
but its presence in the observed ALMA image as well as in our long-term numerical simulation,
suggests its ubiquity.
Here, we have adopted the simulation result at the 73rd orbital period
of the binary, which appears to best resemble the observed 0.9-mm dust-continuum image.

The output from the numerical simulation was transferred to the radiative transfer
calculations, and the theoretically predicted 0.9-mm dust-continuum and C$^{18}$O (3--2)
images were constructed on the assumptions of the LTE condition.
Details of the radiative transfer calculations are also described by Takakuwa et al. (2014).
We then performed CASA observing simulations and created simulated visibility data
for the model images with the same antenna configuration,
hour angle coverage, bandwidth and frequency resolution,
and integration time as those of the real observation.
We further performed flagging of the simulated data to match
the simulated data with the real processed data.
We then made simulated theoretical images with the same imaging methods
as those of the real data.
The weather parameters of the observing simulations were also adjusted to match
the real image noise levels.
In the radiative transfer calculations,
the adopted dust mass opacity and the C$^{18}$O abundance are
$\kappa_{0.9mm}$ = 0.053 cm$^{2}$ g$^{-1}$ and $X_{C^{18}O}$ = 1.7 $\times$ 10$^{-7}$,
respectively.
The temperature profile is adopted to be
$T(r) = \max\left[ 23~{\rm K} \left(\frac{r_A}{300~{\rm AU}}\right)^{-0.2},
19~{\rm K} \left(\frac{r_B}{300~{\rm AU}}\right)^{-0.2} \right]$,
where $r_A$ and $r_B$ indicate the distance from Sources A and B, respectively.
The gas number density at the outermost boundary $R_{bound}$ = 1740 AU
is set to be 1.5 $\times$ 10$^{5}$ cm$^{-3}$.
These parameters were originally derived from our fine tuning to
match the outputs from the radiative transfer
calculations and the observing simulation with the real observed Cycle 0 images.
We confirmed that the same parameters reproduce
the observed Cycle 2 images reasonably well, and thus no further fine tuning of these parameters
was made.

\begin{deluxetable}{ll}
\tablecaption{Parameters for the ALMA Cycle 2 Observation of L1551 NE \label{obs}}
\tabletypesize{\scriptsize}
\tablewidth{0pt}
\tablehead{\colhead{Parameter} &\colhead{Value}}
\startdata
Observing date & 2015 Jun. 27\\
Number of antennas & 36\tablenotemark{a}\\
Field Center & (04$^{h}$31$^{m}$44$\fs$5, 18$\degr$08$\arcmin$31$\farcs$67) (J2000) \\ 
Primary beam HPBW & $\sim$18$\arcsec$ \\
Central Frequency (Continuum) & 335.85 GHz\\
Bandwidth (Continuum) & 1.845 GHz \\
Frequency resolution (C$^{18}$O) & 244.14 kHz $\sim$0.22 km s$^{-1}$\\
Synthesized beam HPBW (Continuum; Natural) & 0$\farcs$190$\times$0$\farcs$169 (P.A. = 178.1$\degr$) \\
Synthesized beam HPBW (Continuum; Uniform) & 0$\farcs$134$\times$0$\farcs$123 (P.A. = 12.3$\degr$) \\
Synthesized beam HPBW (Continuum; $\geq$450 m, Uniform) & 0$\farcs$119$\times$0$\farcs$105 (P.A. = 10.2$\degr$) \\
Synthesized beam HPBW (C$^{18}$O) & 0$\farcs$195$\times$0$\farcs$168 (P.A. = 10.0$\degr$)  \\
Projected baseline coverage & 29 - 1575 m \\
Conversion Factor (Continuum; Natural) & 1 (Jy beam$^{-1}$) = 338.0 (K)\\
Conversion Factor (C$^{18}$O) & 1 (Jy beam$^{-1}$) = 344.4 (K)\\
System temperature &$\sim$100 - 350 K \\
rms noise level (Continuum; Natural)& 0.48 mJy beam$^{-1}$ = 0.16 K\\
rms noise level (C$^{18}$O)& 6.59 mJy beam$^{-1}$ = 2.27 K\\
Flux, Gain calibrator &J0510+180 ($\sim$2.4 Jy)\\
Passband calibrator    &J0423-0120  \\
\enddata
\tablenotetext{a}{For the SO and CS lines as well as the corresponding continuum, the available number of antennas is 35.}
\end{deluxetable}

\section{Results} \label{sec:res}
\subsection{0.9-mm Dust-Continuum Emission} \label{subsec:cont}

\begin{figure}[ht!]
\figurenum{1}
\epsscale{1}
\includegraphics[scale=0.6,angle=0]{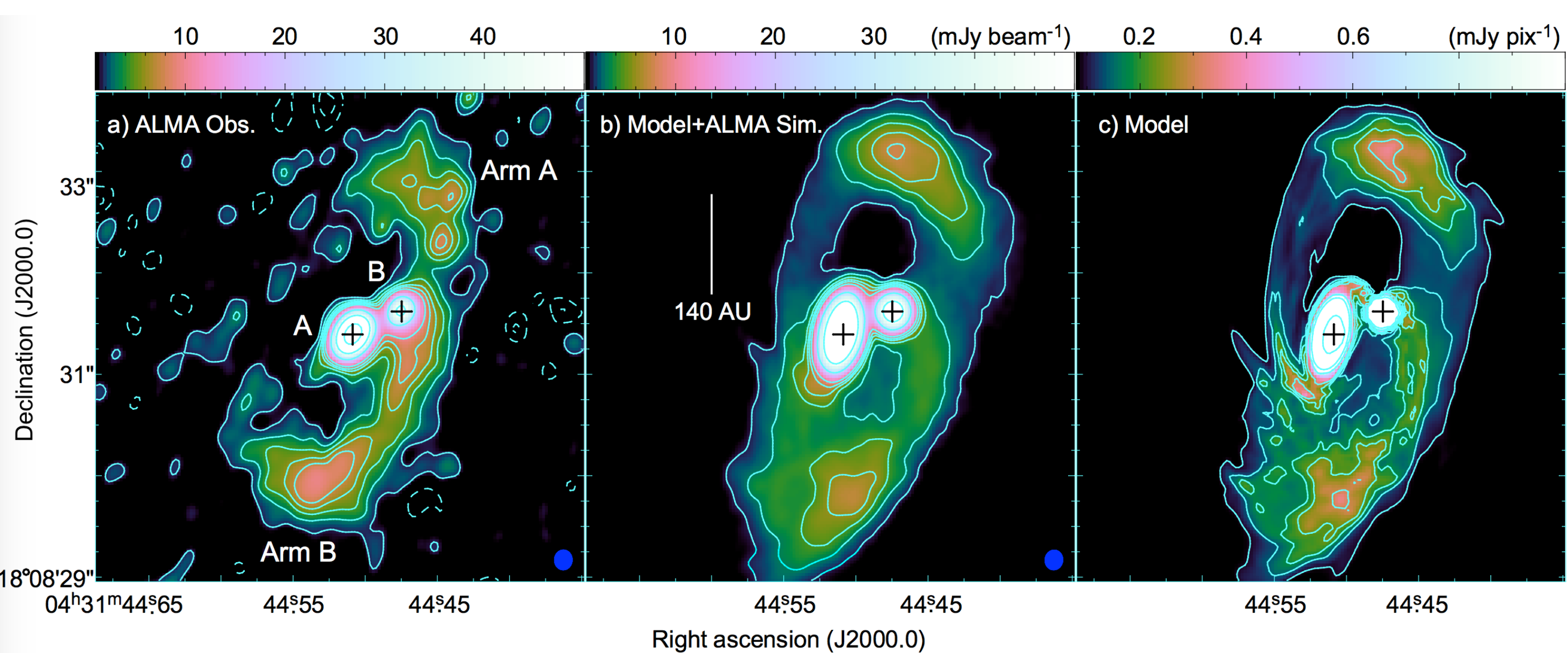}
\caption{a) 0.9-mm dust-continuum image of L1551 NE observed with ALMA.
Contour levels are 3$\sigma$, 6$\sigma$, 9$\sigma$, 12$\sigma$, 15$\sigma$,
20$\sigma$, 50$\sigma$, 100$\sigma$, 200$\sigma$, 400$\sigma$
(1$\sigma$ = 0.48 mJy beam$^{-1}$).
Lower-left and upper-right crosses indicate the centroid positions of
the 2-dimensional Gaussian fittings to the central two dusty components,
which we regard as the positions of Sources A and B.
A filled ellipse at the bottom-right corner shows the synthesized beam
(0$\farcs$190$\times$0$\farcs$169; P.A. = -1.9$\degr$).
b), c) Theoretically-predicted 0.9-mm dust-continuum images of L1551 NE. We performed
the radiative transfer calculation with the gas distribution computed from our 3-D hydrodynamic
model to produce the theoretical image shown in panel c). Then we conducted the ALMA observing
simulation to make the theoretically-predicted ALMA image shown in panel b). Contour levels
in panel b) are the same as those in panel a). Contour levels in panel c) are
15 $\mu$Jy pix$^{-1}$$\times$ 6, 9, 12, 15, 20, 50, 100, 200, 400 (1 pixel = 5 AU$^2$).
\label{fig:cont}}
\end{figure}

Figure \ref{fig:cont}a shows the ALMA Cycle 2 image of L1551 NE
in the 0.9-mm dust-continuum emission with Natural weighting.
There are two intense compact components,
one to the southeast and the other to the northwest.
The configuration of these two components matches well that
of the binary components, Sources A and B, and these components
most likely trace the CSDs around the individual binary members.
From the 2-dimensional Gaussian fitting, the deconvolved sizes of the CSDs
along the major and minor axes ($\equiv D_{maj} \times D_{min}$)
around Sources A and B are measured to be
0$\farcs$29 $\times$ 0$\farcs$19 (40 $\times$ 26 AU) (P.A.$\equiv\theta$=144$\degr$) and
0$\farcs$26 $\times$ 0$\farcs$20 (36 $\times$ 27 AU) ($\theta$=147$\degr$),
respectively.
Since the angular resolution with Natural weighting
is 0$\farcs$190$\times$0$\farcs$169 (P.A. = 178$\degr$),
the CSDs are slightly resolved in the present Cycle 2 observation.
It is thus feasible to discuss the axis ratios
and position angles of the CSDs, whereas with the present beam
it is not possible to study the internal structures of the CSDs.
The corresponding disk inclination angles
($\equiv i = \arccos (\frac{D_{min}}{D_{maj}})$) are calculated
to be 50$\degr$ for Source A and 41$\degr$ for Source B.
To further resolve the CSDs,
we have also made high resolution images in the 0.9-mm dust-continuum emission,
first with Uniform weighting, and then further restricting the visibility data to projected
baseline lengths longer than 450 m
(See Table \ref{obs} for the angular resolutions of those images).
The CSD sizes, inclination and position angles, and the flux densities derived from the 2-dimensional
Gaussian fittings are summarized in Table \ref{disks}.
The results are barely changed with the different
weightings, except for those of Source B at the $>$450 m weighting.
The flux density of the CSD around Source B as measured in the image with
the $>$450 m weighting is considerably ($\sim$factor 2) lower than that
in the other images, suggesting the presence of the severe effect of the missing flux.

On the other hand,
Lim et al. (2016b) have made higher-resolution ($\sim$54 mas = 7.6 AU)
JVLA 7-mm observations of the CSDs in L1551 NE,
which enable them to study the internal structures of the CSDs.
They claimed that the fittings of the NUKER function to the 7-mm image
provide better results than simple 2-dimensional Gaussian fittings,
and derived the breaking diameters, position angles,
and the inclination angles of the Source A disk to be
0$\farcs$266 (37.2 AU), 151$\degr$, and 58$\degr$, and those of the Source B disk
0$\farcs$128 (17.8 AU), 152$\degr$, and 58$\degr$, respectively.
These values are also listed in Table \ref{disks}.
The apparent slight differences of the disk sizes and the inclination angles
between the ALMA and JVLA results
are likely due to the
$\lesssim$3-times lower-angular resolution of our ALMA observation
as compared to the JVLA observation, as well as the differences of the fitting functions
(The errors in the ALMA fitting results only include the statistical errors
of the Gaussian fittings.), and the possible contaminations from the
ionized jet components to the 7-mm emission.
The position angles of the CSD around Source A derived from the ALMA and JVLA observations
are 144$\degr$$\pm$3$\degr$ and and 151$\degr$$\pm$4$\degr$,
and those around Source B are 147$\degr$$\pm$11$\degr$ and and 152$\degr$$\pm$5$\degr$,
respectively. Considering the above-mentioned differences and the statistical errors,
there is no significant difference in the disk position angles between the ALMA
and JVLA results.
The centroid positions of the CSDs derived from the 2-dimensional Gaussian fittings
to the ALMA image are (04$^h$31$^m$44$\fs$509, +18$\degr$08$\arcmin$31$\farcs$396) toward Source A
and (04$^h$31$^m$44$\fs$475, +18$\degr$08$\arcmin$31$\farcs$622) toward Source B.
These positions are within $\sim$50 mas from the corresponding centroid positions of the JVLA image.
The typical astrometric precision of ALMA observations is reported as $\sim$angular resolution / 20,
which is 0$\farcs$18 / 20 $\sim$10 mas in the present observation.
We do not consider, however, that the apparent offsets between the ALMA and JVLA positions
are significant, but they may reflect internal structures of the CSDs, which are not well resolved
with the present ALMA observations. Hereafter in this paper we will adopt the ALMA positions
for Sources A and B.

%
%
%

Our Cycle 2 observations have also revealed detailed structures of the CBD
in L1551 NE. There is an arm-like feature to the south of the protobinary.
The emission ridge of this feature exceeds above 10$\sigma$,
and it clearly shows a curling structure
from the southeast to south, and then to the northwest.
At the northwestern tip, this arm smoothly connects
with the CSD around Source B. Hereafter we call this feature 
``Arm B".
To the north there is another feature (we call this ``Arm A"),
whose emission ridge above 6$\sigma$ also curls from northwest, north
and then northeast.
These two arm-like features comprise the CBD around the protostellar binary.
Compared to our previous lower-resolution 0.9-mm image taken at the Cycle 0 stage,
the new Cycle 2 image presents the curling, arm-like features
more unambiguously, and unveils the connecting feature of Arm B to the CSD of Source B
for the first time.
Furthermore, it is also clear
that the western side of the CBD is much brighter than the eastern side.
Thus, there is a non point-symmetric, $m=1$ mode of the material distribution
in the CBD.
In Figure \ref{fig:cont}b and \ref{fig:cont}c, the simulated theoretical
continuum images after and before the CASA observing simulation are shown
(for the details of the numerical simulation and the radiative transfer calculations, see Takakuwa et al. 2014).
As described above, our time-dependent numerical simulation
shows that after a period of $\sim$20 orbits, such an $m=1$ mode in the CBD
is developed and is then present continuously.
In Figure \ref{fig:cont}b and \ref{fig:cont}c
we show the simulation result of the 73rd orbit.
Our simulation reproduces the observed curling features of Arms A and B,
connection of Arm B to the CSD around Source B, and the brightness distribution
skewed to the west.

\begin{figure}[ht!]
\figurenum{2}
\epsscale{0.5}
\begin{center}
\includegraphics[scale=0.4,angle=0]{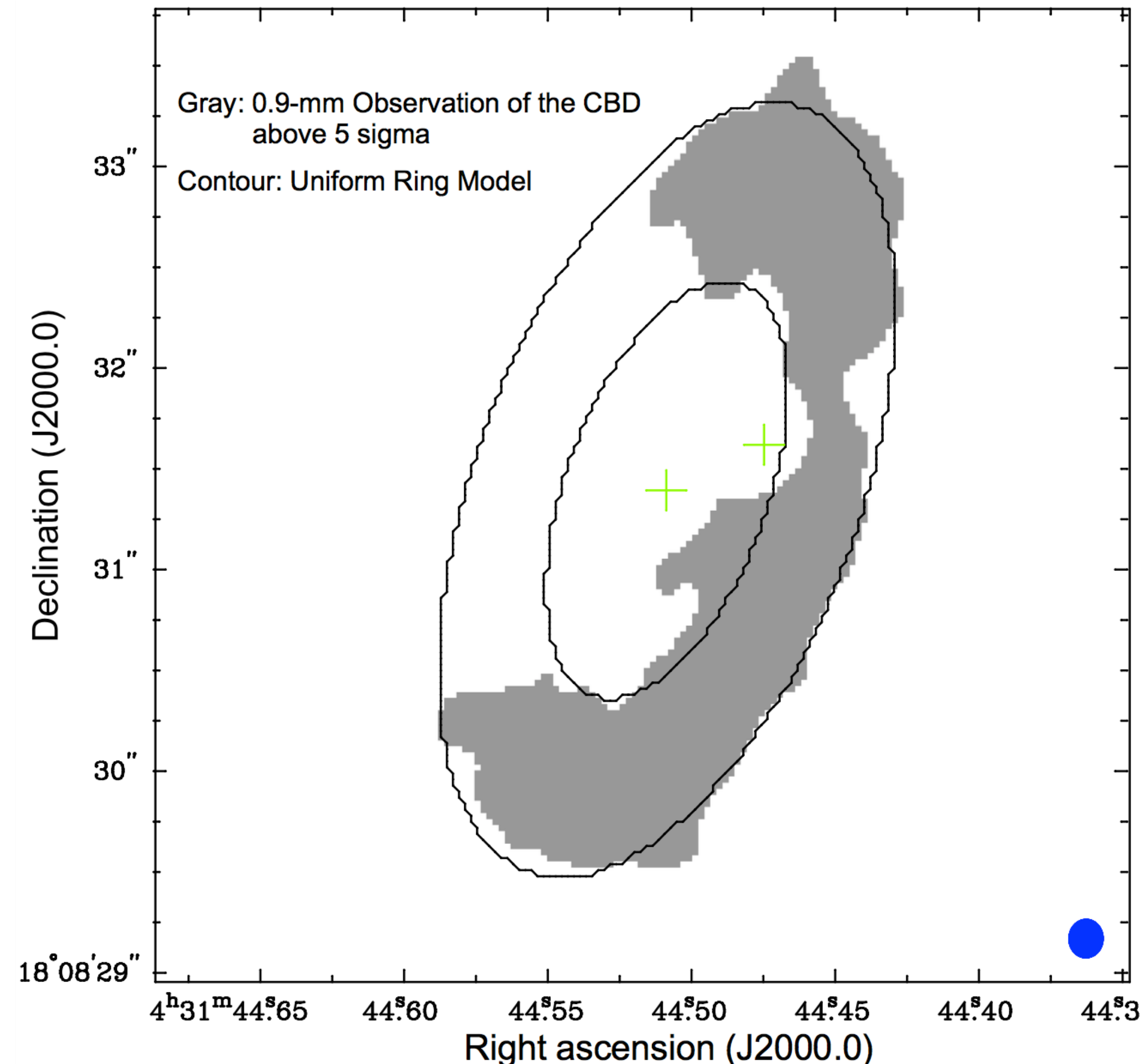}
\end{center}
\caption{Result of the model fitting of an uniform ring to the observed CBD feature,
to estimate the uncertainties of the position and inclination angles of the CBD.
Gray scale denotes the CBD region where the image intensity is higher than 5$\sigma$.
Contours show the best-fit result of the uniform ring with the position and inclination angles
of 160$\degr$ and 62$\degr$, the ring inner diameter of 2$\farcs$2, and the ring width of 0$\farcs$9.
Crosses show the binary positions, and a filled ellipse at the bottom-right corner is 
the synthesized beam.
\label{fig:ringmodelfit}}
\end{figure}

From kinematical model fitting of Keplerian rotation to
the C$^{18}$O (3--2) velocity channel maps of L1551 NE observed with
the SMA, the position and inclination angles of the CBD have been estimated to be
$\theta$ = 167$\degr^{+23\degr}_{-27\degr}$ and $i$ = 62$\degr^{+17\degr}_{-25\degr}$,
respectively \cite{tak12}.
These values are adopted in our simulation, which reproduces
the observed features of the CBD (Figure \ref{fig:cont}).
From the 0.9-mm dust-continuum image taken with the ALMA Cycle 2 observation,
we have also attempted to re-estimate $\theta$ and $i$ and their uncertainties.
Since the structure of the CBD is complicated it is not straightforward to make
a fitting of any simple geometrical shape (such as a 2-dimensional Gaussian)
to the observed CBD image and to derive the fitting uncertainties.
We have thus made the following simplification.
First, the CSD components as derived from the 2-dimensional Gaussian fittings described above
were subtracted from the observed 0.9-mm image and the CBD-only image was generated.
Then, inside and outside the regions with the image intensities higher than 5$\sigma$,
the intensities were replaced with an uniform intensity of unity and zero, respectively.
We then regarded the simplified structure of the CBD as an uniform circular ring,
and performed $\chi^2$ model fitting with $\theta$, $i$, the ring inner diameter,
and the ring width as the fitting parameters.
The fitting result is shown in Figure \ref{fig:ringmodelfit}.
The estimated position and inclination angles are
$\theta$ = 160$\degr^{+12\degr}_{-14\degr}$
and $i$ = 62$\degr^{+9\degr}_{-8\degr}$, consistent with our canonical values
adopted in our numerical simulation.
This simple ring fitting shows that the typical uncertainties of $\theta$ and $i$ are
$\pm13\degr$ and $\pm9\degr$, respectively.

The masses of the CSDs and CBD ($\equiv M_{d}$) are estimated
from their individual continuum fluxes ($\equiv S_{\nu}$) measured in the
Naturally-weighted image as,
\begin{equation}
M_{d}=\frac{S_{\nu}d^2}{\kappa_{\nu} B_{\nu}(T_d)},
\end{equation}
where $\nu$ is the frequency, $d$ the distance, $B_{\nu}(T_d)$ the Planck function
for dust at a temperature $T_{d}$, and $\kappa_{\nu}$ the dust opacity per unit gas + dust mass
on the assumption of a gas-to-dust mass ratio of 100.
The same dust mass opacity at 0.9-mm as that adopted in Takakuwa et al. (2014),
$\kappa_{0.9mm}$ = 0.053 cm$^{2}$ g$^{-1}$, is used, which is calculated
from $\kappa_{\nu}$ = $\kappa_{\nu_{0}}$($\nu$/$\nu_{0}$)$^{\beta}$,
$\kappa_{250~\mu m}$=0.1 cm$^{2}$ g$^{-1}$ \cite{hil83}, and $\beta$=0.5 \cite{gui11,chi12}.
The same range of the dust temperature as that adopted in Takakuwa et al. (2014),
$T_d$ = 10 - 42 K, is adopted (For comparison, the peak brightness temperature
of the C$^{18}$O (3--2) image cube ranges $\sim$10--35 K.).
Then, a mass for the CSD of Source A is
calculated to be $\sim$0.005 - 0.043 $M_{\odot}$, that of Source B
$\sim$0.002 - 0.016 $M_{\odot}$, and that of the CBD
$\sim$0.009 - 0.076 $M_{\odot}$. These mass estimates are in close agreement
with those from our previous Cycle 0 observation \cite{tak14}.
The adopted dust mass opacity ($\kappa_{0.9mm}$ = 0.053 cm$^{2}$ g$^{-1}$)
is a factor 3 higher than that of Ossenkopf \& Henning (1994) for grains
with thin ice mantles coagulated at a density of 10$^{6}$ cm$^{-3}$
($\kappa_{0.9mm}$ = 0.018 cm$^{2}$ g$^{-1}$). Thus, adopting the dust mass opacity
by Ossenkopf \& Henning (1994) provides a factor 3 higher masses of the CSDs and CBD.
The mass ranges derived above do not take this uncertainty of the dust mass opacity
into account. In any case,
the CSD and CBD masses are much smaller than the inferred
total binary mass of 0.8 $M_{\odot}$, suggesting that the self-gravity of
the disks is negligible compared to the gravitational field of the binary protostars.

\begin{deluxetable}{lcccccc}
\tablecaption{Circumbinary and Circumstellar Disk Parameters
in L1551 NE \label{disks}}
\tabletypesize{\scriptsize}
\tablewidth{0pt}
\tablehead{\colhead{Disk} &\colhead{Weighting} &\colhead{$D_{maj}$$\times$$D_{min}$}
&\colhead{$i$} &\colhead{$\theta$} &\colhead{$S_{\nu}$} &\colhead{$M_{d}$\tablenotemark{a}} \\
\colhead{} &\colhead{} &\colhead{(AU$\times$AU)}
&\colhead{($\degr$)} &\colhead{($\degr$)} &\colhead{(Jy)} &\colhead{($\times$10$^{-3}$ $M_{\odot}$)}}
\startdata
Source A CSD      &Natural  &(40$\pm$1)$\times$(26$\pm$1) &50$^{+2}_{-3}$ &144$\pm$3 &0.34 &5.0--43.2 \\
                  &Uniform  &(40$\pm$1)$\times$(26$\pm$1) &49$^{+2}_{-2}$ &145$\pm$2 &0.33 &4.9--42.3 \\
                  &$>$450 m\tablenotemark{b} &(39$\pm$1)$\times$(22$\pm$1) &56$^{+2}_{-2}$ &145$\pm$2 &0.25 &3.7--32.0 \\
Source A CSD 7-mm\tablenotemark{c} &Natural  &37$\times$20 &58 &151$\pm$4 &\nodata &\nodata \\           
Source B CSD &Natural  &(36$\pm$2)$\times$(27$\pm$2) &41$^{+8}_{-11}$ &147$\pm$11 &0.13 &1.9--15.9 \\
             &Uniform  &(31$\pm$2)$\times$(24$\pm$2) &40$^{+8}_{-10}$ &143$\pm$12 &0.11 &1.6--13.6 \\
             &$>$450 m\tablenotemark{b} &(22$\pm$1)$\times$(13$\pm$1) &52$^{+6}_{-8}$ &139$\pm$9 &0.06 &0.9--7.6 \\
Source B CSD 7-mm\tablenotemark{c} &Natural  &18$\times$9 &58 &152$\pm$5 &\nodata &\nodata \\           
CBD          &Natural &$\sim$600$\times$280\tablenotemark{d} &$\sim$62$^{+9}_{-8}$\tablenotemark{e} &$\sim$167$^{+5}_{-21}$\tablenotemark{e} &0.60\tablenotemark{f} &8.8--76.0\tablenotemark{f} \\
\enddata
\tablenotetext{a}{Assuming $T_d$=10--42 K and $\kappa_{0.9 mm}$=0.053 cm$^2$ g$^{-1}$ . See texts for details.}
\tablenotetext{b}{Uniform weighting of the visibility data with the projected baseline lengths longer than 450 m.}
\tablenotetext{c}{From the fitting of the JVLA 7-mm image to the NUKER function by Lim et al. (2016b).}
\tablenotetext{d}{From the approximate zeroth intensity level in Figure \ref{fig:cont}.}
\tablenotetext{e}{See texts for details.}
\tablenotetext{f}{Flux density after the subtraction of the CSD components from the 2-dimensional Gaussian fitting
in Figure \ref{fig:cont}.}
\end{deluxetable}


\subsection{Distributions of the Molecular-Line Emission in L1551 NE} \label{subsec:mol}

\begin{figure}[ht!]
\figurenum{3}
\epsscale{0.7}
\begin{center}
\includegraphics[scale=0.5,angle=0]{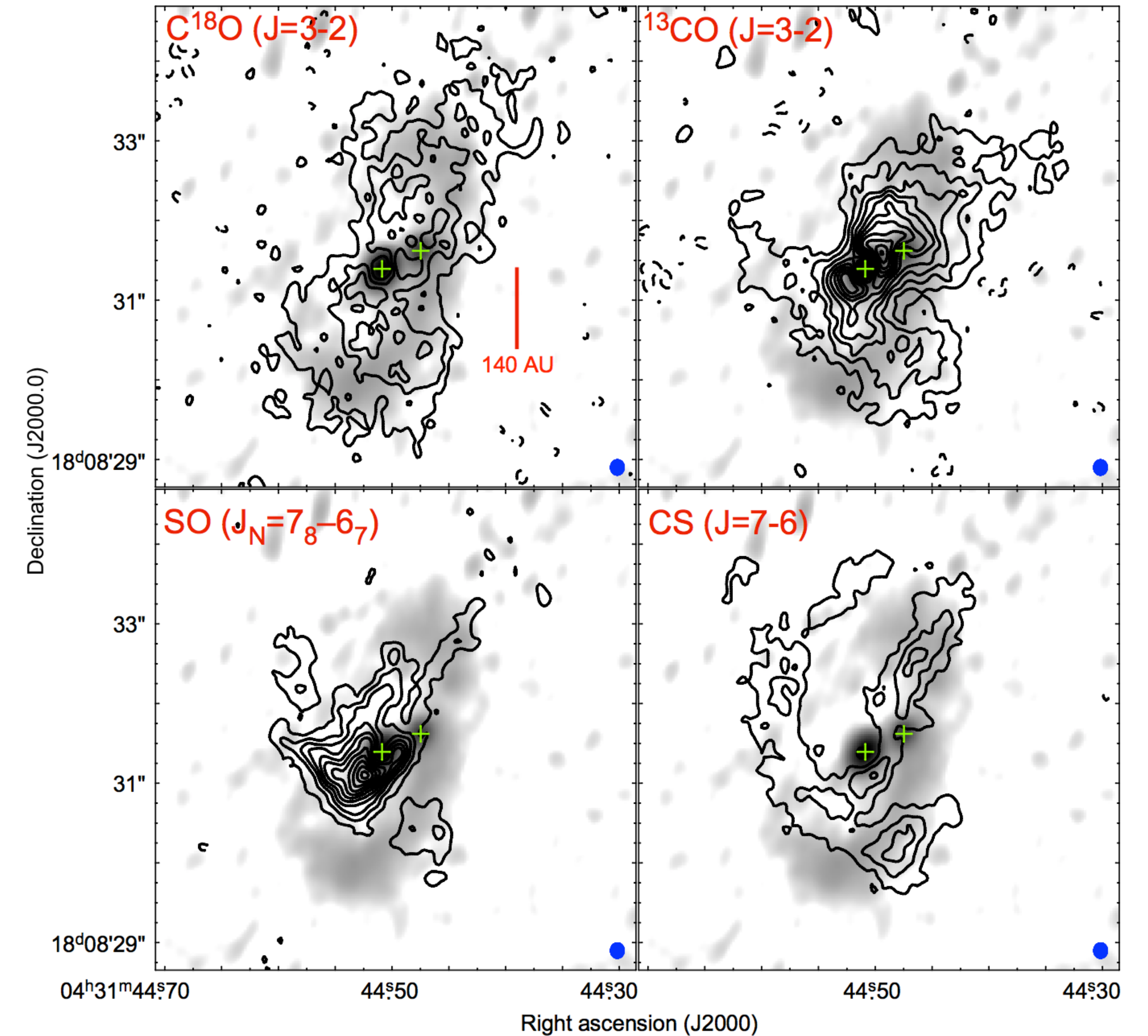}
\end{center}
\caption{Moment 0 maps of the observed molecular lines as labeled (contours),
superposed on the 0.9-mm continuum image (gray; same as Figure \ref{fig:cont}a), in L1551 NE.
Contour intervals are 5$\sigma$, where
the 1$\sigma$ levels are  6.71 mJy km s$^{-1}$ (= 2.31 K km s$^{-1}$),
7.58 mJy km s$^{-1}$ (= 2.60 K km s$^{-1}$),
6.85 mJy km s$^{-1}$ (= 2.25 K km s$^{-1}$), and
6.85 mJy km s$^{-1}$ (= 2.29 K km s$^{-1}$) in the C$^{18}$O, $^{13}$CO, SO, and CS maps, respectively.
Integrated velocity ranges are 2.18--11.51 km s$^{-1}$, -0.11--12.95 km s$^{-1}$,
1.36--12.10 km s$^{-1}$, and 2.28--9.53 km s$^{-1}$
in the C$^{18}$O, $^{13}$CO, SO, and CS maps, respectively.
\label{fig:lines}}
\end{figure}

Figure \ref{fig:lines} shows the total integrated intensity (moment 0) maps
of the C$^{18}$O ($J$=3-2), $^{13}$CO ($J$=3-2),
SO ($J_N$=7$_8$-6$_7$), and the CS ($J$=7-6) lines (contours),
superposed on the 0.9-mm dust-continuum image (gray scale).
These four lines exhibit different spatial distributions.
While the overall emission extent and the elongation
of the C$^{18}$O emission are consistent with those of the continuum emission,
the $^{13}$CO emission distribution is different from the C$^{18}$O distribution,
and is more concentrated inside the arm-like features of the 0.9-mm dust-continuum emission.
The SO emission is seen predominantly in the vicinity of Source A,
and its peak brightness temperature is $>$46 K
to the southeast of Source A. Recent ALMA studies of protostars in
the SO (6$_5$-5$_4$; 219.949 GHz) line,
found that the SO emission traces shocks associated with the accretion onto
the central disks from the surrounding protostellar envelopes \cite{sak14,yen14,oha14}.
It is thus possible that the intense SO emission
in the vicinity of Source A also traces shocks associated with the accretion from the CBD to
the CSD around Source A.
The CS (7--6) emission exhibits a tilted $E$-shaped feature,
and it is not easy to correlate the CS emission distribution with
the dust-emission distribution.
These different molecular-line distributions
imply significant chemical inhomogeneities in the CBD around L1551 NE.

Since the C$^{18}$O emission appears to trace the CBD structure as seen in the 0.9-mm
dust-continuum emission most straightforwardly, and since the velocity resolutions of the
$^{13}$CO and the SO, CS data are a factor of 2 and 4 worse, respectively,
in the following we will use the C$^{18}$O line to discuss the velocity structures
of the CBD and CSDs in L1551 NE.

%

\subsection{C$^{18}$O (3--2) Emission} \label{subsec:c18o}

\begin{figure}[ht!]
\figurenum{4}
\epsscale{1.3}
\begin{center}
\includegraphics[scale=0.8,angle=0]{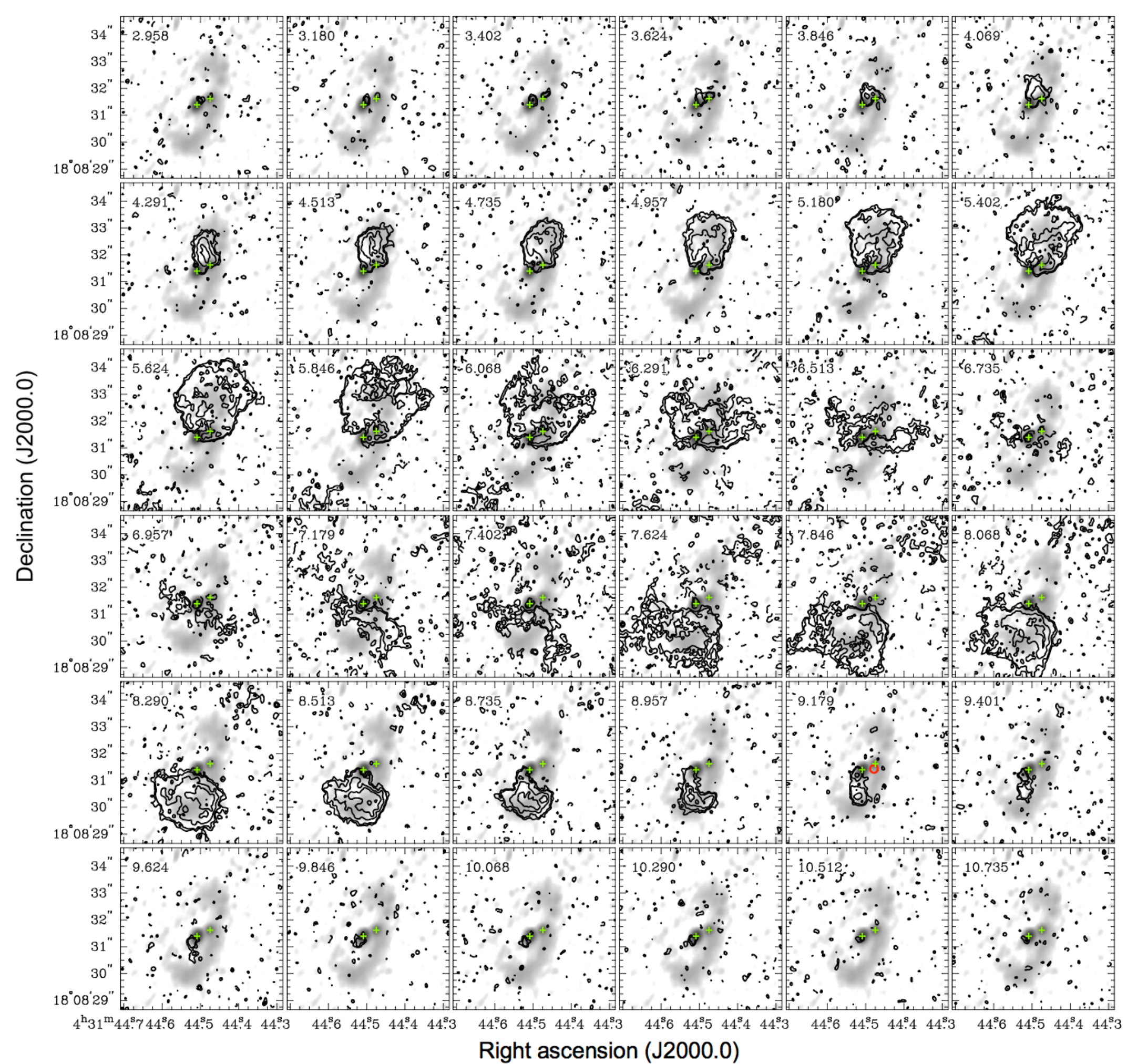}
\end{center}
\caption{Velocity channel maps of the C$^{18}$O (3--2) emission (contours),
superposed on the 0.9-mm continuum image (gray; same as Figure \ref{fig:cont}a),
in L1551 NE.
Contour levels are 3$\sigma$, 5$\sigma$, and then in steps of 5$\sigma$
(1$\sigma$ = 4.66 mJy). A red open circle at the velocity of 9.2 km s$^{-1}$
denotes the anticipated location of
the redshifted counterpart of the high-velocity blueshifted emission to
the northwest of Source B (see texts for details).
\label{fig:c18och}}
\end{figure}

Figure \ref{fig:c18och} shows the velocity channel maps of the C$^{18}$O (3--2) line
in contours, superposed on the 0.9-mm dust-continuum image of L1551 NE in gray scale.
Our previous SMA and ALMA Cycle 0 observations of L1551 NE in the C$^{18}$O (3--2) line
have found that the systemic velocity of the CBD is
$V_{\rm LSR}$=6.9 km s$^{-1}$ \cite{tak12,tak13,tak14,tak15}.
Centered on this systemic velocity, our ALMA Cycle 2 observation of L1551 NE detected
the C$^{18}$O emission from $V_{\rm LSR}$=3.0 to 10.7 km s$^{-1}$.
At the most blueshifted velocities, two compact C$^{18}$O
emission components, one to the northwest of Source A and the other northwest of Source B, are seen.
To the southeast of Source A the redshifted counterpart is also seen at the most
redshifted velocity, while such a redshifted
counterpart is not seen to the southeast of Source B.
At slightly lower velocities, the blueshifted C$^{18}$O emission extends northeast
from the midpoint between Sources A and B, while the redshifted C$^{18}$O emission
extends to the south and then southwest. The blueshifted C$^{18}$O emission then
expands to the western side too, and traces the entire northern side of the CBD.
A similar tendency is also seen in
the redshifted side, where
the C$^{18}$O emission extends to the east and traces the entire southern half of the CBD.
Around the systemic velocity
a characteristic ``butterfly'' pattern, a signature of a Keplerian disk \cite{sim00},
is seen.

\begin{figure}[ht!]
\figurenum{5}
\epsscale{1.0}
\begin{center}
\includegraphics[scale=0.6,angle=0]{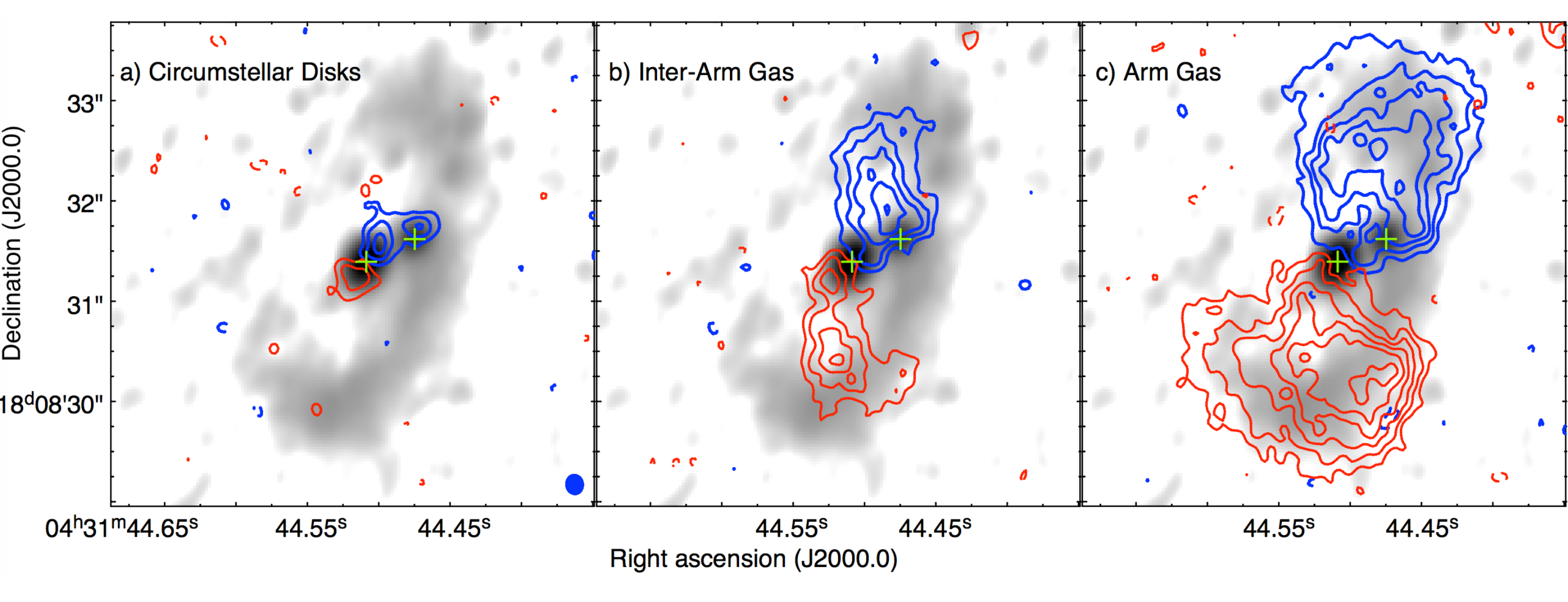}
\end{center}
\caption{Maps of the blueshifted (blue contours) and redshifted (red) C$^{18}$O (3--2) emission
at representative velocities, superposed on the 0.9-mm continuum image
(gray scale; same as Figure \ref{fig:cont}a), in L1551 NE.
Integrated velocity ranges are 2.18--3.96 km s$^{-1}$
(blueshifted) and 9.62--11.40 km s$^{-1}$ (redshifted) in panel a, 3.96--4.51 km s$^{-1}$
and 8.85--9.62 km s$^{-1}$ in panel b, and 4.51--5.29 km s$^{-1}$ and 8.18--8.85 km s$^{-1}$
in panel c. Contour levels are in steps of 5$\sigma$
(1$\sigma$ = 2.93 mJy km s$^{-1}$ in panel a, 1.64 and 1.94 mJy km s$^{-1}$
for the blueshifted and redshifted emission in panel b, and 1.94 and 1.79 mJy km s$^{-1}$
for the blueshifted and redshifted emission in panel c).
\label{fig:c18o3}}
\end{figure}

Figure \ref{fig:c18o3} represents three primary components identified
in the C$^{18}$O maps; circumstellar-disk components, inter-arm gas, and
arm-gas components. In Figure \ref{fig:c18o3}a, the blueshifted and
redshifted C$^{18}$O components are located
symmetrically with respect to Source A, and these C$^{18}$O components
most likely trace the CSD as seen in the 0.9-mm dust continuum emission
around Source A. In Figure \ref{fig:c18o3}b, the blueshifted C$^{18}$O emission
extends northward from the midpoint between Sources A and B, and the redshifted emission extends
southward from the southeast of Source A. The tip of the blueshifted emission reaches to
the eastern part of Arm A, and that of the redshifted emission reaches to the western part of Arm B.
In Figure \ref{fig:c18o3}c, the blueshifted and redshifted C$^{18}$O emission
trace the structures of Arms A and B, respectively.

In Figure \ref{fig:c18o3}b, it is clear that to the east of Source A the C$^{18}$O
emission is redshifted and to the west blueshifted.
Since the eastern and western sides of the CBD are near and far-side, respectively,
the detected east (redshifted)-west (blueshifted) velocity gradient is consistent
with an infalling gas motion toward Source A.
In Figure \ref{fig:c18o3}a,
another blueshifted C$^{18}$O component is seen
to the northwest of Source B, however, there is no redshifted counterpart
seen to the southeast of Source B. In the $^{13}$CO (3--2) image cube,
this redshifted counterpart in Source B, is not seen either above the 2$\sigma$ noise level
(1$\sigma \sim$4.5 mJy),
while the other components are clearly seen. These results imply
that the blueshifted component to the northwest of Source B does not trace
the CSD.


%
%
%

Below, we will discuss the C$^{18}$O emission components associated with the CSDs and the CBD
separately.

\subsubsection{CSDs} \label{subsubsec:csds}

From our previous SMA and ALMA studies of L1551 NE, the total binary mass, mass ratio,
and the inclination and position angles of the CBD are estimated to be
0.8$M_{\odot}$, 0.19, 62$\degr$ and 167$\degr$, respectively.
Then, on the assumption of the circular Keplerian orbit of the binary system
co-planar to the CBD, the line-of-sight velocities of Sources A and B
are calculated to be $V_{LSR}$ = 7.0 km s$^{-1}$ and 6.3 km s$^{-1}$, respectively.
The high-velocity blueshifted and redshifted C$^{18}$O emission around Source A
(Figure \ref{fig:c18o3}a)
are located approximately symmetrically with this systemic velocity of Source A.
In the case of Source B,
the high-velocity blueshifted emission is seen around
$V_{LSR} \sim$ 3.4 km s$^{-1}$, and thus the redshifted counterpart should appear
around $\sim$9.2 km s$^{-1}$.
Such a redshifted component is, however, not seen in the C$^{18}$O and
$^{13}$CO image cubes. This lack of emission cannot be explained
by the contamination from the CO emission in the CBD, as
there is no CBD CO emission at the relevant location and velocity
(see a red open circle in Figure \ref{fig:c18och}).

\begin{figure}[ht!]
\figurenum{6}
\epsscale{1.0}
\begin{center}
\includegraphics[scale=0.6,angle=0]{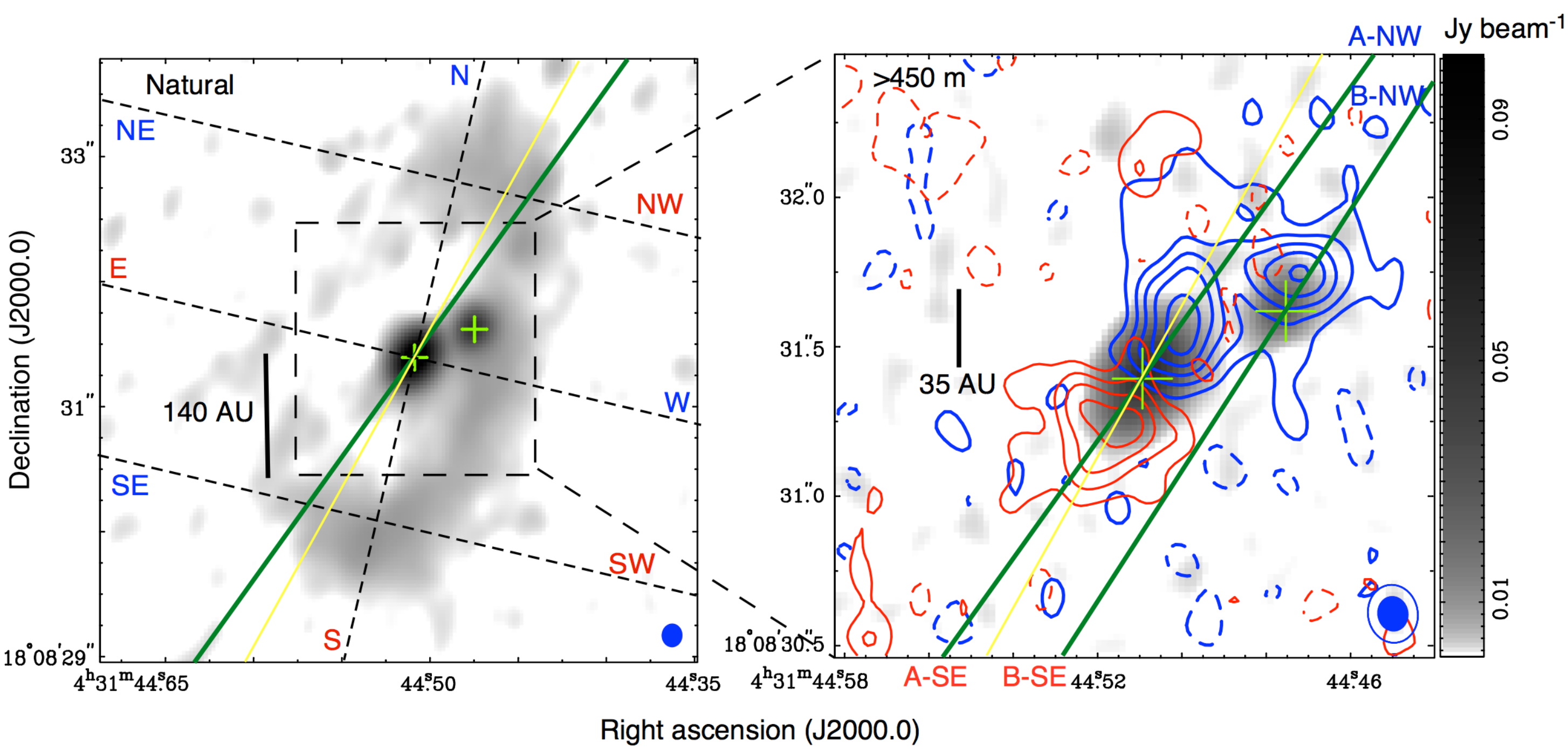}
\end{center}
\caption{Comparison between the CBD and the CSDs in L1551 NE.
The left panel shows the naturally-weighted
0.9-mm dust-continuum image of L1551 NE (same as Figure \ref{fig:cont}a),
$i.e.$, the entire image of the CBD.
The right panel shows a zoom-up view of the CSDs around Sources A and B. Gray
scales in the right panel depicts the 0.9-mm dust-continuum image made from the visibility
data with the projected baseline lengths longer than 450 m.
The blue and red contours show the
high-velocity blueshifted and redshifted C$^{18}$O (3--2) emission (same
as those in Figure \ref{fig:c18o3}a).
Filled and open ellipses at the bottom-right corner denote the
beams of the continuum and C$^{18}$O images, respectively (see Table \ref{obs}).
Green lines passing through Source A in the left and right panels show
the direction of the major axis of the 0.9-mm dust-continuum emission
associated with Source A.
Another green line passing through Source B in the right panel shows
the major axis of the 0.9-mm dust-continuum emission around Source B.
Yellow lines show the direction of the major axis of
the 7-mm continuum emission of Source A as observed with JVLA \cite{lim16b}.
Dashed lines with N,S,..,SW denote the cut lines
of the Position-Velocity (P-V) diagrams shown in Figures \ref{fig:pv1} and \ref{fig:pv2}.
Green lines in the right panel with A-NW etc show the P-V cut lines of Figure \ref{fig:pvA}. 
\label{fig:misalign}}
\end{figure}

Figure \ref{fig:misalign} (right) overlays the distribution of the
high-velocity C$^{18}$O emission (contours) on the
higher-resolution 0.9-mm dust-continuum image (gray scale). In this continuum image
emission structures originated from the CBD are resolved out, but
the angular resolution is a factor $\sim$1.6 better than that
with the Natural weighting (see Table \ref{obs}). Thus,
the elongations of the CSDs as seen in the 0.9-mm dust-continuum emission
are better resolved.
The position angle of the CSD around Source A as measured from this higher-resolution
dust-continuum image (=145$\degr$$\pm$2$\degr$; see Green lines in Figure \ref{fig:misalign})
matches well with the axis connecting the blueshifted and redshifted C$^{18}$O emission.
These high-velocity C$^{18}$O emission components associated with the CSD have been unveiled for the first time
with the present higher-resolution and high-sensitivity Cycle 2 observation.
Furthermore, the CSD major axis appears to be tilted with respect
to the major axis of the CBD (Figure \ref{fig:misalign} left)
\footnote{In our numerical simulations the position and inclination angles of the CSDs are
assumed to be the same as those of the CBD.}.
From our simplified uniform ring model fitting to the CBD image as described
in section \ref{subsec:cont}, the position angle of the CBD and its uncertainty
is estimated to be $\theta$ = 160$\degr^{+12\degr}_{-14\degr}$. While the lower
bound of $\theta$ of the CBD and the upper bound of $\theta$ of the CSD around Source A
slightly overlap, the probability to match the both values is only 
$\sim (\frac{1-0.6826}{2})^2  = 2.5 \%$. In addition, the estimated inclination angle
of the CBD and that of the CSDs around Sources A and B are also different by more than their
uncertainties (see Table \ref{disks}). These results imply that the CBD and the CSDs
are misaligned and not co-planar with each other.


%

\begin{figure}[ht!]
\figurenum{7}
\epsscale{0.6}
\begin{center}
\includegraphics[scale=0.8,angle=0]{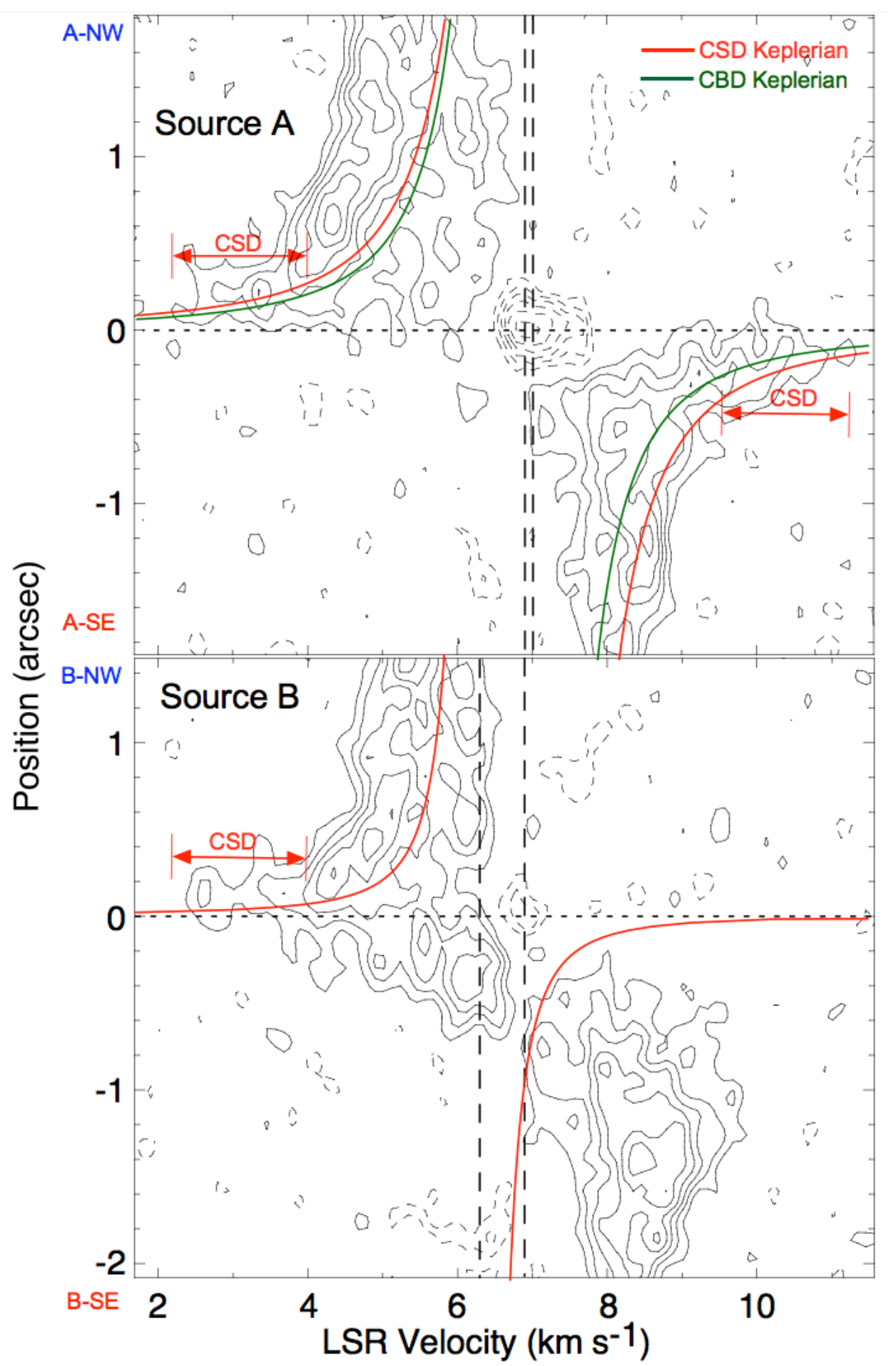}
\end{center}
\caption{P-V diagrams of the
C$^{18}$O ($J$=3-2) emission in L1551 NE along the major axes of the CSDs around
Sources A (upper panel; P.A.=144$\degr$) and B (lower panel; P.A.=147$\degr$)
(see Figure \ref{fig:misalign}).
Contour levels are in steps of 2$\sigma$ (1$\sigma$ = 6.59 mJy beam$^{-1}$).
Horizontal dashed lines in the upper and lower panels denote
the positions of Sources A and B, respectively.
A common vertical dashed line in the upper and lower panels
shows the systemic velocity of the CBD (= 6.9 km s$^{-1}$). A right-hand dashed line
in the upper panel denotes the inferred light-of-sight velocity
or the systemic velocity of Source A (= 7.0 km s$^{-1}$),
and a left-hand dashed line in the lower panel the systemic velocity
of Source B (= 6.3 km s$^{-1}$).
Green curves in the upper panel denote the Keplerian rotation curve of the CBD.
Red curves in the upper panel show the inferred Keplerian rotation curves
of the CSD around Source A, where the disk inclination angle and the central stellar mass
are -50$\degr$ and 0.675 $M_{\odot}$, respectively. Red curves in the lower panel
show the Keplerian rotation curve of Source B,
where the disk inclination angle and the central stellar mass
are -41$\degr$ and 0.125 $M_{\odot}$, respectively.
Because the PV cut line is not along the major axis of the CBD, the Keplerian
rotation curve of the CBD shows lower velocities than that of the CSD around Source A.
Red horizontal arrows show the velocity regions of the CSDs around Sources A and B.
\label{fig:pvA}}
\end{figure}

Figure \ref{fig:pvA} shows the Position-Velocity (P-V) diagrams of the C$^{18}$O emission
along the major axes of the CSDs of Sources A (upper panel) and B (lower panel) 
(see Figure \ref{fig:misalign}). While the bulk of the C$^{18}$O emission arises
from the CBD, the C$^{18}$O emission originated from the CSDs can be identified
as emission protrusions at the highest velocities (red arrows in Figure \ref{fig:pvA}).
Red curves in the upper and lower panels of
Figure \ref{fig:pvA} shows the inferred Keplerian rotation curves
of the CSDs around Sources A and B, respectively.
%
In the P-V diagram of Source A, the emission ridge originated from
the CSD appears to trace the relevant Keplerian rotation curve.
On the blueshifted side, the emission peak is located at the positional offset of $\sim$0$\farcs$21
and $V_{LSR}$ $\sim$3.4 km s$^{-1}$. At the same positional offset the Keplerian rotation curve
yields $V_{LSR}$ = 3.5 km s$^{-1}$. On the redshifted side, the offset position of the emission
peak at $V_{LSR}$ = 9.8 km s$^{-1}$ is $\sim$0$\farcs$28, and at the same positional offset
the Keplerian rotation curve predicts $V_{LSR}$ = 10.0 km s$^{-1}$. The differences
of the velocities are less than the observational velocity resolution ($\sim$0.22 km s$^{-1}$).
As shown in Figure \ref{fig:misalign},
the direction of the velocity gradient of the C$^{18}$O emission
is along the major axis of the CSD around Source A.
These results imply that the high-velocity C$^{18}$O emission associated
with Source A traces the Keplerian rotation of the CSD around Source A.
On the contrary, the ridge of the high-velocity blueshifted emission
to the northwest of Source B deviates from the inferred Keplerian rotation curve.
There are two peaks in the emission ridge; one at $V_{LSR}$ = 2.6 km s$^{-1}$
and the other at $V_{LSR}$ = 3.7 km s$^{-1}$, both of which have a similar
positional offset $\sim$0$\farcs$13. At the same positional offset
the Keplerian rotation curve predicts $V_{LSR}$ = 4.7 km s$^{-1}$,
and hence the observed peak velocities are $\gtrsim$1.0 km s$^{-1}$ higher.
This implies that the velocity of the blueshifted C$^{18}$O emission around Source B
is too high for the Keplerian rotation. In addition, the redshifted counterpart
is not seen.
These results suggest that the origin of the high-velocity blueshifted
C$^{18}$O emission to the northwest of Source B is unlikely to be the Keplerian rotation.
Instead, this high-velocity blueshifted emission may trace a gas component flowing into
the Roche lobe.

\subsubsection{CBD} \label{subsubsec:cbd}

\begin{figure}[ht!]
\figurenum{8}
\epsscale{0.6}
\begin{center}
\includegraphics[scale=0.6,angle=0]{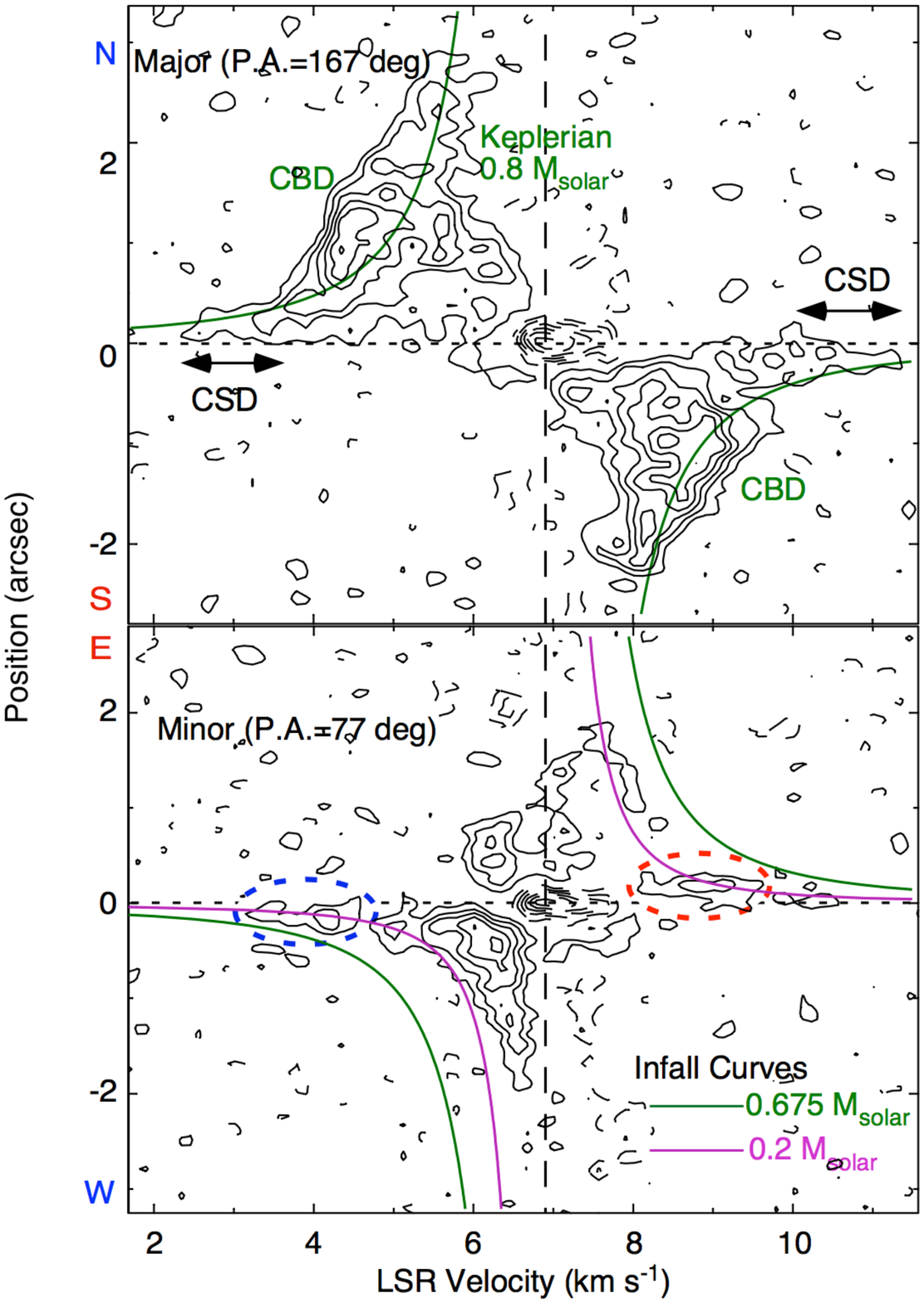}
\end{center}
\caption{P-V diagrams of the
C$^{18}$O ($J$=3-2) emission in L1551 NE along the major (P.A.=167$\degr$; $upper$ $panel$)
and minor axes (P.A.=77$\degr$; $lower$) of the CBD, passing through the position of Source A
(see Figure \ref{fig:misalign} left.).
Contour levels are in steps of 2$\sigma$ (1$\sigma$ = 6.59 mJy beam$^{-1}$).
Horizontal and vertical dashed lines denote the position of Source A and
the systemic velocity 6.9 km s$^{-1}$, respectively.
A green curve in the upper panel shows the Keplerian rotation curve
with the central stellar mass of 0.8 $M_{\odot}$ and the
disk inclination angle of 62$\degr$. Arrows indicate the velocity regions
of the CSD around Source A. Colored curves in the lower panel
denote curves of free-fall with the central stellar masses as labeled.
Blue and red dashed ellipses highlight the detected infalling gas components
toward Source A.
\label{fig:pv1}}
\end{figure}

Figure \ref{fig:pv1} shows P-V diagrams of the C$^{18}$O emission
along the major and minor axes of the CBD. The central position of these P-Vs
is adopted to be the position of Source A.
Since the inferred binary mass ratio ($\sim$0.19) is small,
the exact dynamical center of the CBD is only $\sim$0$\farcs$08 northwest from
the Source A (= primary) position.
We have confirmed that this
small offset of the central position does not produce any recognizable change
of the P-V diagrams. Thus, for simplicity, hereafter we will adopt the P-Vs
passing though the position of Source A, to discuss the velocity structures of the CBD.

In the P-V diagram along the major axis, the bulk of the C$^{18}$O emission traces the rotating
CBD, and the spatial-velocity distribution is consistent with the Keplerian rotation curve
on the assumption of the central stellar mass of 0.8 $M_{\odot}$ and
the disk inclination angle of 62$\degr$.
At the highest velocities, emission components originated from the CSD of Source A
are also present (arrows in Figure \ref{fig:pv1}).
In the P-V diagram along the minor axis, there are low-velocity
blueshifted ($V_{LSR} \sim$5.0-6.7 km s$^{-1}$) components to the east and west
of Source A. These components trace the outer CBD components, including the ``butterfly"
emission components (see Figure \ref{fig:c18och}).
In addition to these low-velocity blueshifted components,
there are high-velocity blueshifted 
and redshifted
components in the close vicinity ($\lesssim 0\farcs2$) of Source A (dashed ellipses in Figure \ref{fig:pv1}).
The high-velocity blueshifted
and redshifted components are located to the west and east of Source A,
respectively, and the sense of the velocity gradient is consistent with
the infalling motion toward Source A.
The peak of the high-velocity blueshifted
emission resides at the positional offset of -0$\farcs$11 and
$V_{LSR}$ = 4.1 km s$^{-1}$, and that of the high-velocity redshifted emission
at +0$\farcs$19 and $V_{LSR}$ = 9.1 km s$^{-1}$.
If the identified infall motion is free-fall toward the central stellar mass
of 0.675 $M_{\odot}$ ($i.e.,$ the inferred mass of Source A), at the positional
offset of the blueshifted component,
the line of sight velocity should be $V_{LSR}$ = 1.4 km s$^{-1}$, which is
much higher than the observed velocity (see green curves in Figure \ref{fig:pv1}).
Similarly, at the positional offset
of the redshifted component the velocity should be $V_{LSR}$ = 11.0 km s$^{-1}$,
too high as compared to the observed velocity. Instead, if the central stellar
mass is 0.2 $M_{\odot}$ (red curves in Figure \ref{fig:pv1}), at the positional
offsets of the blueshifted and redshifted emission peaks the velocities are
calculated to be 3.9 and 9.1 km s$^{-1}$, consistent with the observed velocities.
These results indicate that the velocities of the identified infalling motion are
smaller than those of free-fall.

\begin{figure}[ht!]
\figurenum{9}
\epsscale{0.8}
\begin{center}
\includegraphics[scale=0.5,angle=0]{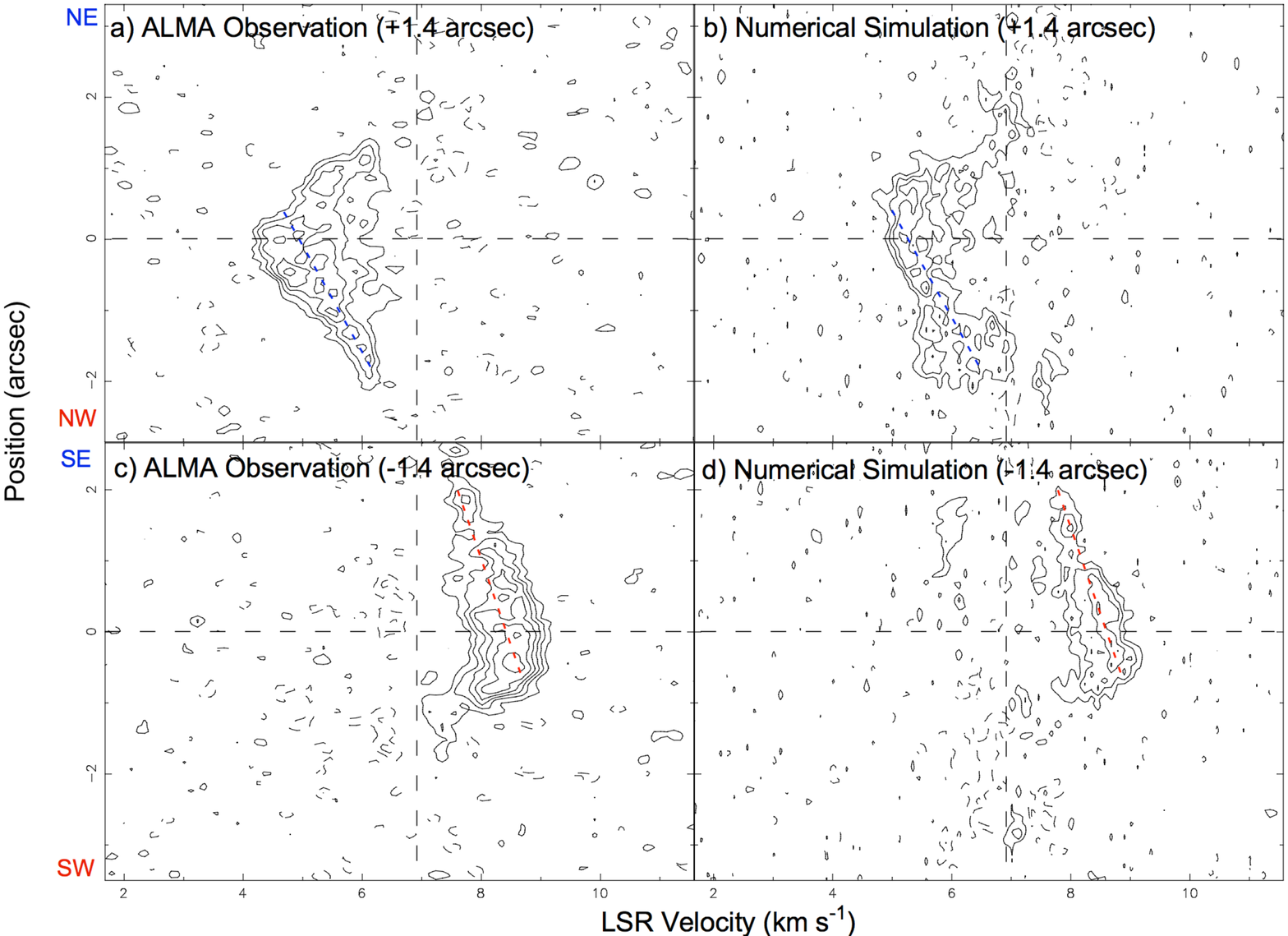}
\end{center}
\caption{Observed (left panels) and model (right) P-V diagrams of the
C$^{18}$O ($J$=3-2) emission in L1551 NE along the cuts parallel to the minor axis
of the CBD passing through $\pm$1.4$\arcsec$ offsets from the position of Source A
along the major axis (see Figure \ref{fig:misalign} left).
Contour levels are in steps of 2$\sigma$ (1$\sigma$ = 6.59 mJy beam$^{-1}$).
Horizontal and vertical dashed lines denote the position of Source A
and the systemic velocity 6.9 km s$^{-1}$, respectively.
Dashed lines denote the velocity gradients detected along the off-center minor axes.
\label{fig:pv2}}
\end{figure}

Figure \ref{fig:pv2} (left) shows the observed P-V diagrams of the C$^{18}$O emission
along the transverse directions of Arms A and B (NE-NW and SE-SW in Figure \ref{fig:misalign} left).
Across Arm A, there is a velocity gradient where
the eastern part is more blueshifted than the western part
(blue dashed line in Figure \ref{fig:pv2}). A similar velocity gradient
is also seen across Arm B (red dashed line).
Since the eastern part of the CBD is on the near-side and the western part is on the far-side,
the identified velocity gradients across Arms A and B suggest presence of expanding
gas motions in the arms,
as already found with the previous Cycle 0 observation \cite{tak14}.

\section{Discussion} \label{sec:dis}
\subsection{Kinematics of the CBD} \label{subsec:grav}

Our ALMA Cycle 2 observation of L1551 NE has resolved the CBD in the 0.9-mm
dust-continuum emission into two arm-like features, Arms A and B, where Arm B is connected
to the CSD around Source B. Furthermore, the western part of the CBD is much brighter
than the eastern part, suggesting the presence of the $m=1$ mode of the material distribution.
These observed continuum features are reproduced with our numerical simulation of the CBD.

\begin{figure}[ht!]
\figurenum{10}
\epsscale{1.2}
\begin{center}
\includegraphics[scale=0.6,angle=0]{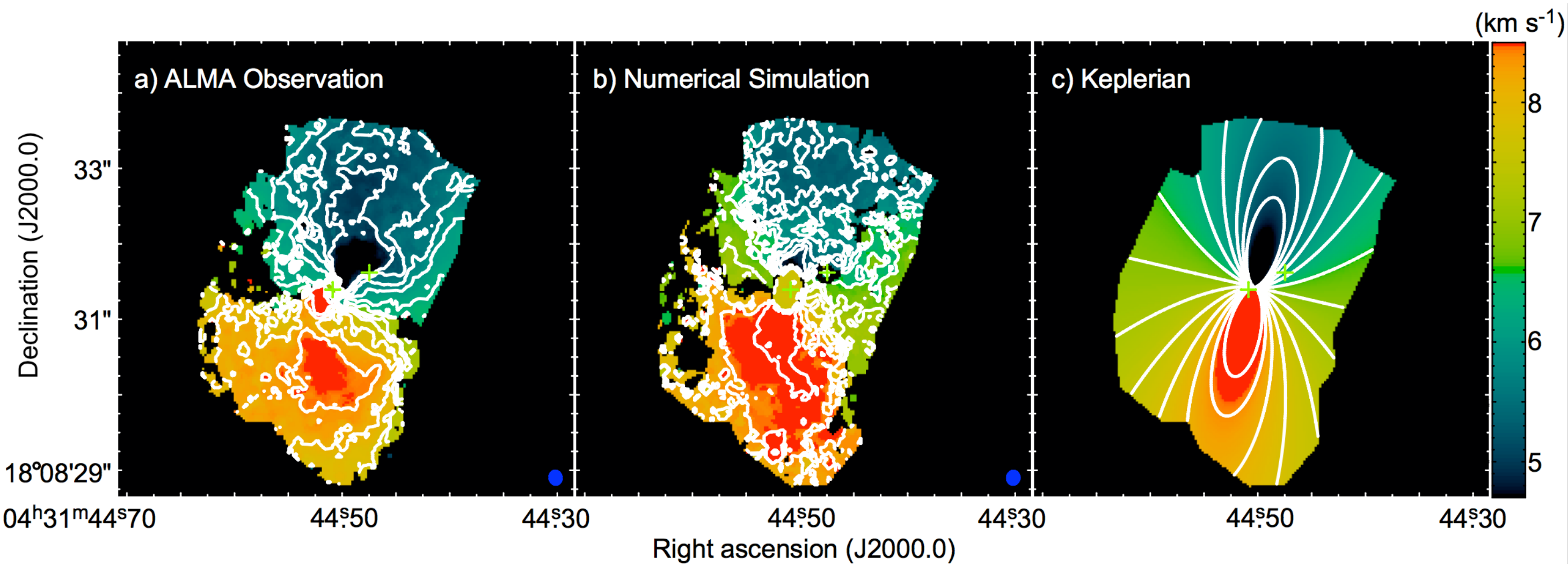}
\end{center}
\caption{Moment 1 maps of the observed C$^{18}$O emission (a), our numerical model (b), and
the Keplerian-disk model with the central stellar mass of 0.8 $M_{\odot}$ (c), in L1551 NE.
The contour interval is 0.3 km s$^{-1}$, and the bluest and reddest contour
levels are 5.1 km s$^{-1}$ and 8.7 km s$^{-1}$, respectively.
\label{fig:mom1}}
\end{figure}

In the C$^{18}$O line the global rotation of the CBD and the expanding gas motions
in Arms A and B have been identified. Furthermore,
the infalling gas components in the vicinity ($r \lesssim 50$ AU) of Source A have
also been found.
Figure \ref{fig:mom1} compares the observed C$^{18}$O mean-velocity map, simulated
C$^{18}$O mean-velocity map, and the anticipated mean-velocity map of the
circular Keplerian rotation with the central stellar mass of 0.8$M_{\odot}$.
In the case of the circular Keplerian rotation the mean velocity map should be symmetric
with respect to the major axis. The observed and simulated mean-velocity
maps show, however, slight deviations from the Keplerian rotation.
In these maps the southwestern part is
slightly more redshifted than the southeastern part,
and the northeastern part is slightly more blueshifted than the northwestern part.
These deviations from Keplerian reflect expanding gas motions in the arms.
Indeed, the simulated P-V diagrams across Arms A and B shown in Figure \ref{fig:pv2} (right)
reproduce the expanding motions in the arms found with our ALMA observation.
We note that in the radiative transfer calculation, artificially high dust-opacity
``masks" with a radius of 20 AU have been put at the locations of Sources A and B,
to hide the two ``sinks" adopted in our numerical simulation.
Therefore, the observed infalling components in the vicinity of Source A
cannot be reproduced with our model.

%


Our numerical simulation, as well as previous
theoretical studies of CBDs around protostellar binaries
\cite{bat00,gun02,och05,han10,you15}, predict that the non-axisymmetric
gravitational torques from the binary, induce both inward and outward gas
motions with slower and faster rotations than the Keplerian rotations,
respectively, and creates spiral density patterns in the CBDs.
At the radii outside the L2 and L3 Lagrangian points,
spiral arms in the CBDs are the regions where the gravitational torques
impart angular momenta and drive
faster rotation than the Keplerian rotation. As a result
the spiral regions expand radially. On the other hand,
the gravitational torques extract angular momenta from the inter-arm
regions, and thus drive infall.
The observed expanding gas motions in Arms A and B, and infall
in the inter-arm regions, can thus be interpreted as a result of
the non-axisymmetric gravitational torques of the binary.

Our Cycle 2 observations have also unveiled that the material distributions in the CBD around
L1551 NE, are not point-symmetric, but skewed to the western side
(Figure \ref{fig:cont}a).
The presence of such an $m=1$ mode is also reproduced by our numerical simulation
as described above.
The asymmetric structure of the CBD rotates
in the direction of the orbital motion of the binary stars, and its angular
velocity is lower than that of the orbital motions. The asymmetry therefore
rotates with respect to the position angles of the binary stars.
The adopted 73rd orbital period of the simulation, gives
one of the best configurations of the asymmetry and the binary position angle,
and shapes of the arms (Figures \ref{fig:cont}b and \ref{fig:cont}c.).
The theoretical discussion on the asymmetry of the CBD will be presented in
a forthcoming paper.

\subsection{Infall onto the Protostellar Binary} \label{subsec:inf}

As described in section \ref{subsubsec:cbd}, an infalling motion toward Source A,
whose velocity is slower than the free-fall velocity,
is observed in the CBD.
The infalling motion in the CBD could determine the final binary mass and
the mass ratio, as discussed in previous theoretical studies
\cite{bat00,gun02,och05,han10,you15}.
%
The mass of the infalling component ($\equiv M_r$) is estimated by integrating the blueshifted
and redshifted C$^{18}$O emission
in the vicinity of Source A from Figure \ref{fig:c18o3}b, on the assumption of
the excitation temperature of the C$^{18}$O emission of 42 K \cite{bar93,mor94},
$X$(C$^{18}$O)=1.7$\times$10$^{-7}$ \cite{cra04}, and the LTE condition.
The mass of the infalling component is estimated to be
$M_r$ $\sim$8.1$\times$10$^{-5}$ $M_{\odot}$. Then, the mass infalling rate ($\equiv \dot{M}$)
can be estimated as
\begin{equation}
\dot{M}=\frac{M_r v_r}{r},
\end{equation}
and the value is $\dot{M} \sim$1.0$\times$10$^{-6}$ $M_{\odot}~yr^{-1}$.
%
From $\dot{M}$ the accretion luminosity ($\equiv L_{acc}$) can also be calculated as
\begin{equation}
L_{acc}=\frac{G M_{\star} \dot{M}}{R_{in}},
\end{equation}
where $G$ is the gravitational constant, $M_{\star}$ is a mass of the central protostar,
and $R_{in}$ is the terminal radius of the infall.
If we assume that the infalling gas directly reaches to the surface of the protostar
(i.e., $R_{in}$ = $R_{\star}$ = 4$R_{\odot}$) \cite{sta80}, the accretion luminosity is $\sim$5.4 $L_{\odot}$, similar to the bolometric luminosity of L1551 NE (= 4.2 $L_{\odot}$; Froebrich 2005).

In the above estimates, we simply extracted the C$^{18}$O emission
with the east (red) to west (blue) velocity gradient in the vicinity of Source A,
since the velocity gradient along the minor axis can be interpreted as an infalling motion.
On the other hand, it is not straightforward to observationally identify infalling motions
along the disk major axis, because the radial motion along the major axis
does not have any velocity vector along the line of sight (LOS).
Thus, the mass infalling rate estimated above should be regarded
as the lower limit of the entire infall toward Source A.
Toward Source B, our ALMA observations do not identify infalling motions (i.e., a velocity
gradient along the minor axis centered on Source B).
However, this does not indicate that there is no infall toward Source B,
because infalling gas motions perpendicular to the observational line of sight
can not be measured.
Previous studies of infall in protostellar envelopes assume axisymmetric
gas motions to estimate the entire mass infalling rate from the
observed velocity gradient along the minor axis \cite{mom98,tak07,yen10,yen11,tak13}.
On the other hand,
in the vicinity of protostellar binaries axisymmetric gas motions cannot be assumed
because the gravitational field is non-axisymmetric. Thus it is much more difficult
to observationally estimate the infalling rates, and compare the mass infalling rates
onto the primary and secondary.
Our numerical simulation of L1551 NE indeed shows highly asymmetric distributions
of the radial (infall and expansion) motions (see Figure 9c in Takakuwa et al. 2014)
as well as the azimuthal motions (Figure 9b).
Detailed comparisons of our numerical simulations and the observational image cubes
will be the subject to our forthcoming paper.

In addition to mass infall in CBDs onto the protostellar binaries, kinematics of surrounding
protostellar envelopes which can replenish CBDs with fresh materials should also be
taken into account. Our previous ASTE observations of the large-scale ($\sim$20000 AU)
protostellar envelope in L1551 NE have unveiled that the protostellar envelope
is being dispersed by the redshifted outflow driven from the neighboring protostar,
L1551 IRS 5 \cite{tak15}. While in the inner $\lesssim$1000 AU scale there is an infalling
component \cite{tak13}, the mass of that component ($\sim$0.002 $M_{\odot}$)
is smaller than the CBD mass (0.009 -- 0.076 $M_{\odot}$; Table \ref{disks}).
Thus, further mass supply from the protostellar envelope
to the CBD in L1551 NE is not expected.
The inferred present masses of Sources A and B
are $\sim$0.68 $M_{\odot}$ and $\sim$0.13 $M_{\odot}$, respectively,
and even if all the amount of the material in the CBD plus the infalling component
is accreted onto the binary, the binary mass and the mass ratio do not change much.
Therefore, even though L1551 NE is a Class I protostellar binary associated
with the CBD and the protostellar envelope, the mass and the mass ratio are
already close to the final values.

\subsection{Misalignments of the CSDs from the CBD in L1551 NE} \label{subsec:misa}

Our high-resolution Cycle 2 observations of L1551 NE have resolved the elongation of the
dust components around the individual binary protostars,
and measured the position and inclination angles of the CSDs unambiguously.
Furthermore, the high-velocity blueshifted and redshifted C$^{18}$O emission
are associated with the CSD of Source A, and
the major axis of the CSD
matches with the direction of the velocity gradient.
The radii and velocities
of the C$^{18}$O emission are consistent with the Keplerian rotation
with the inferred Source A mass of 0.675 $M_{\odot}$ (Figure \ref{fig:pvA}).

As discussed in sections \ref{subsec:cont} and \ref{subsubsec:csds},
the position and inclination angles of the CSDs around Sources A and B
are different from those of the CBD.
These results indicate that the CSDs
are misaligned from the CBD, and that the CSDs are not co-planar with the CBD.
Such a non-coplanar configuration could cause precessing motions of the CSDs, which are
lighter than the CBD, on the time scale of the binary orbital period. The precessing
amplitude must be on the order of the difference of the position angles between the
CBD and the CSD ($\sim$20$\degr$).
Previous NIR observations have found that Source A drives collimated jets along
the northeast (red) to southwest (blue) direction \cite{dev99,rei00,mor06,hay09}.
The jets consists of a chain of knots of HP2 (HH 454A), HH29 and 28 on the
blueshifted side, and HP3, 4 (=HH 454B), 5, and 6 on the redshifted side.
All of these knots are within 2$\degr$ of a common jet axis at P.A. $\sim$61$\degr$
(The P.A. of the minor axes of the CSD of Source A and the CBD are 54$\degr$
and 77$\degr$, respectively.).
The proper motions of these HH knots are also measured to be $\sim$150 km s$^{-1}$,
and the dynamical timescale of the most distant HH knot, HH 28, is measured to be
$\sim$2700 yr. For comparison, the orbital period of the Sources A and B binary
system is calculated to be $\sim$1700 yr, on the assumption of the circular orbit.
Thus, for the last $\sim$1.6 orbital periods there is no significant (within
the P.A. difference between the jet and the CSD minor axis of 61$\degr$-54$\degr$ = 7$\degr$)
precession in the disk-jet system of Source A,
and the misaligned configuration appears to be rather static.
%
%
%
%
%
%

In the presence of the misalignment between the CBD and CSDs, the infalling material
and the angular momenta from the CBD do not smoothly connect to
the outermost radii of the CSDs.
The infalling materials should fall onto the disk surface or even onto the protostar directly.
Indeed, from the argument in section \ref{subsec:inf}
direct infall from the CBD onto the protostar is required to account for the protostellar
luminosity with the accretion luminosity.
Furthermore, if the infalling materials fall onto the disk surface, such direct impacts
could affect the physical and chemical conditions of the CSD significantly,
and the subsequent disk evolution into the planet formation stage.
As shown in Figure \ref{fig:lines}, the SO (7$_8$--6$_7$) emission is concentrated
around Source A, and the peak brightness temperature of the SO emission is
$>$46 K to the southeast of Source A.
In L1489 IRS \cite{yen14} and L1527 IRS \cite{sak14,oha14},
SO lines have been observed in the transitional regions from
the infalling envelopes to the inner Keplerian disks.
Such SO emission distributions can be interpreted as
the presence of the accretion shocks,
although in the other protostars the interpretation of the SO emission
distributions is not straightforward \cite{yen17}.
It is thus possible that the observed intense SO emission
in the close vicinity of Source A may also trace shocks associated with the impacts
of the infalling materials from the CBD onto the CSD surface.
Such SO emission distribution
associated with the accretion onto CSDs from the CBD has indeed
been found in another binary system of UY Aurigae \cite{tan14}.

\section{Summary} \label{sec:sum}

We have performed the ALMA Cycle 2 observations of the protostellar binary system
L1551 NE in the 0.9-mm dust-continuum,
C$^{18}$O ($J$=3-2), $^{13}$CO ($J$=3-2), SO ($J_N$=7$_8$-6$_7$), and the CS ($J$=7-6) emission
at an angular resolution of $\sim$0$\farcs$18, four times higher than that
of our previous Cycle 0 observations.
The present Cycle 2 observations have unveiled the detailed structures and kinematics
of the CBD in L1551 NE, and resolved the CSDs around
the individual binary protostars. We have also performed numerical simulations
and constructed the theoretically-predicted 0.9-mm and C$^{18}$O images of L1551 NE, in order 
to interpret the observed structures and gas motions in the CBD.
The main results are summarized below.

\begin{itemize}
\item[1.] The 0.9-mm dust-continuum emission traces two CSDs around
the individual binary protostars, and the CBD in L1551 NE.
The CSDs are spatially resolved and the deconvolved sizes, position and
inclination angles of the CSD around Source A are measured to be
40 $\times$ 26 AU, $\theta = 144\degr \pm 3\degr$, and $i = 50\degr^{+2\degr}_{-3\degr}$,
and those around Source B are
36 $\times$ 27 AU, $\theta$ = 147$\degr \pm 11\degr$, and $i = 41\degr^{+8\degr}_{-11\degr}$.
The position and inclination angles of the CSD around Source A are different
from those of the CBD ($\theta = 160\degr^{+12\degr}_{-14\degr}$ and $i = 62\degr^{+9\degr}_{-8\degr}$),
and the inclination angle of the CSD around Source B
are also different from that of the CBD.
The CBD exhibits two arm-like features (Arm A to the north and Arm B to the south),
and Arm B connects to the CSD of Source B. Furthermore, the brightness distribution
of the continuum emission is skewed to the western part of the CBD, suggesting the presence
of the $m=1$ mode of the material distribution in the CBD.
The masses of the CSDs around Sources A and B, and the CBD are estimated to be
0.005-0.043 $M_{\odot}$, 0.002-0.016 $M_{\odot}$, and 0.009--0.076 $M_{\odot}$
for $T_d$ = 10--42 K, respectively.

\item[2.] The high-velocity blueshifted (3.0-3.6 km s$^{-1}$)
C$^{18}$O emission is associated with the northwestern halves of the CSDs
as seen in the 0.9-mm dust-continuum emission. The redshifted
(9.6-10.7 km s$^{-1}$) counterpart of the C$^{18}$O emission is also seen
in the southeastern half of the CSD of Source A, whereas such a redshifted
counterpart is not found toward Source B. In Source A,
the direction of the velocity gradient of the high-velocity blueshifted
and redshifted C$^{18}$O emission matches well with the major axis of the CSD.
In the intermediate velocity ranges
the blueshifted C$^{18}$O emission (3.9-4.5 km s$^{-1}$)
connects between the midpoint between Sources A and B
and the eastern part of Arm A, and the redshifted C$^{18}$O emission
(9.0-9.4 km s$^{-1}$) between the southeast of Source A
and the western part of Arm B.
There is also a east (red)-west (blue) velocity gradient
centered on Source A.
In the lower velocity range the blueshifted emission extends to the western part of Arm A
and the redshifted emission to the east of Arm B. This suggests the presence of
east (blue)-west (red) velocity gradients across the arms, and the sense of those
velocity gradients is opposite to that around Source A.

\item[3.] The east (red)-west (blue) velocity gradient around Source A in the CBD represents
infalling motion toward Source A, whereas the opposite velocity gradients
seen across Arms A and B represent expanding motion in the arms. The two arm structures and the expanding motions, and the infall in the
inter-arm regions, can all be reproduced with our numerical simulation.
Our numerical simulation predicts that the non-axisymmetric gravitational torques
of the binary, impart angular momenta to the spiral arms in the CBD
and drive faster rotation than the Keplerian rotation and expansion,
while the gravitational torques extract angular momenta from the inter-arm
regions and thus drive infall.
Furthermore, our numerical simulation also predicts the $m=1$ mode of
the material distribution in the CBD.

\item[4.] The spatial-velocity distributions of the high-velocity C$^{18}$O emission
associated with the CSD around Source A are consistent with the Keplerian rotations with the inferred
central stellar masses (=0.68 $M_{\odot}$). 
Furthermore, the measured differences of the position and inclination angles of the
CSDs around Sources A and B from those of the CBD
imply that the CSDs are misaligned with respect to the plane of the CBD.
As the axis of the NIR jet knots driven from Source A is close to orthogonal to
the major axis of the CSD and appears unchanged
over the past $\sim$1.6 orbital period of the binary system (1 orbital period $\sim$1700 yr),
the misalignment of the CSD is rather static.
Since the CSDs are likely misaligned from the CBD, the infalling materials
and the angular momenta from the CBD do not smoothly connect to
the outermost radii of the CSDs, but directly falls onto the surfaces of the CSDs.
Such impacts of the infalling material could significantly change the physical and chemical
condition of the CSDs. The concentrated SO emission in the vicinity of Source A
may reflect shocks associated with the impact of the infalling material.

\end{itemize}


\acknowledgments
We would like to thank N. Ohashi, M. Hayashi, and M. Momose for their
fruitful discussions, and all the ALMA staff supporting this work.
S.T. acknowledges a grant from the Ministry of Science and Technology (MOST) of Taiwan
(MOST 102-2119-M-001-012-MY3), and JSPS KAKENHI Grant Number JP16H07086,
in support of this work.
J.L. is supported by the GRF grants of the Government of the Hong Kong SAR under HKU
703512P for conducting this research. T.M. is supported by the
Grants-in-Aid for Scientific Research (26400233) from the Ministry of Education,
Culture, Sports, Science and Technology, Japan.
This paper makes use of the following ALMA data:
ADS/JAO.ALMA\#2013.1.00349.S. ALMA is a partnership of ESO (representing
its member states), NSF (USA) and NINS (Japan), together with NRC (Canada)
and NSC and ASIAA (Taiwan) and KASI (Republic of Korea), in
cooperation with the Republic of Chile. The Joint ALMA Observatory is
operated by ESO, AUI/NRAO and NAOJ.
Numerical computations were in part carried out on Cray XC30 at
Center for Computational Astrophysics, National Astronomical Observatory of Japan.



\vspace{5mm}
\facilities{ALMA}

\software{CASA, Miriad}

\listofchanges

\end{document}